**High-throughput computational characterization of two-dimensional compositionally complex transition-metal chalcogenide alloys**

Duo Wang,[1] Lei Liu,[1] Neha Basu,[1,2] and Houlong L. Zhuang[1]*

[1]School for Engineering of Matter, Transport and Energy, Arizona State University, Tempe, AZ 85287, USA

[2]BASIS Scottsdale High School, Scottsdale, AZ 85259, USA

*zhuanghl@asu.edu

**Abstract**

Two-dimensional (2D) binary transition-metal chalcogenides (TMCs) like molybdenum disulfide exhibits excellent properties as materials for light adsorption devices. Alloying binary TMCs can form 2D compositionally complex TMC alloys (CCTMCAs) that possess remarkable properties from the constituent TMCs. We adopt a high-throughput workflow performing density functional theory (DFT) calculations based on the virtual crystal approximation (VCA) model (VCA-DFT). We test the workflow by predicting properties including in-plane lattice constants, band gaps, effective masses, spin-orbit coupling (SOC), and band alignments of the Mo-W-S-Se, Mo-W-S-Te, and Mo-W-Se-Te 2D CCTMCAs. We validate the VCA-DFT results by computing the same properties using unit cells and supercells of selected compositions. The VCA-DFT results of the abovementioned five properties are comparable to that of DFT calculations, with some inaccuracies in several properties of MoSTe and WSTe. Moreover, 2D CCTMCAs can form type II heterostructures as used in photovoltaics. Finally, we use $Mo_{0.5}W_{0.5}SSe$, $Mo_{0.5}W_{0.5}STe$, and $Mo_{0.5}W_{0.5}SeTe$ 2D CCTMCAs to demonstrate the room-temperature entropy-stabilized alloys. They also exhibit high electrical conductivities at 300K, promising for light adsorption devices. Our work shows that the high-throughput workflow using VCA-DFT calculations provides a tradeoff between efficiency and accuracy, opening up opportunities in the computational design of other 2D CCTMCAs for various applications.

**Introduction**

Binary two dimensional (2D) transition metal chalcogenides (TMCs) exhibit strong in-plane chemical bonds and weak out-of-plane interactions, allowing for stable monolayers [1,2]. The properties of 2D TMCs such as $MX_2$ have been investigated from both computational simulation and the experimental works [3-7]. Strain engineering on $MX_2$ gives rise to tunable properties such as band gaps and effective masses. Moreover, monolayer $MX_2$ have direct band gaps ranging from 1.0 to 2.0 eV [3]. They also display thickness-dependent electronic properties of band gap [8], which transitions from indirect in multiple layers and bulk $MX_2$ to direct in monolayer $MX_2$. Owing to these excellent properties, binary 2D TMCs have recently grown in prominence with their promising applications such as photovoltaics, photodetectors, and field-effect transistors [4,7,9,10].

However, a general problem of binary 2D TMCs, is their fixed properties such as band gaps, which limit their applications in fields such as optoelectronics and photovoltaics where adjustable band gaps are needed to accommodate different wavelength ranges [11-14]. To maximize the efficiency in light-electricity energy conversion, different modification processes such as alloying [9,10], mechanical straining [15], and forming heterostructures [16] have been applied for achieving tunable electronic properties such as band gaps. One of these methods is to obtain van der Waals (vdW) heterostructures by stacking together two different binary monolayers [17]. Different vdW heterostructures have many applications such as field-effect tunneling transistors [18], photovoltaics [19], and other optoelectronic devices [20]. 2D TMCs based vdW heterostructures have been used in heterojunction photovoltaics. For example, the $WS_2/WSe_2$ vdW heterostructure, has been investigated via DFT calculations and experiment and shown to be a promising candidate for photovoltaics because of the high light adsorption efficiency and high carrier mobility [21,22]. In addition to vertical vdW heterostructures, lateral heterostructures from two or more TMCs have recently also been predicted by theoretical calculations [23-25] and demonstrated in experiments [22,26].

Apart from forming heterostructures, alloying multiple materials of different elements together to form bulk compositionally complex alloys (CCAs) such as CoCrFeMnNi and Al$_x$CoCrFeNi [27-30] has been shown to allow for a high tunability in band structures. In contrast to conventional alloys that are comprised of one or two principal elements and of much lower percentages of other elements [31,32], CCAs encompass not only conventional alloys but also high-entropy alloys that have more than five principal elements of equal or near-equal molar ratios [33-35]. In CCAs, the composition consisting of multiple elements has fundamental effects on configurational entropy, free energy, phase selection, and stability [33,36]. For example, the increasing of temperature will cause the decrease of Gibbs free energy in the system of high configurational entropy, thus increasing the stability of CCAs [37]. Moreover, despite the complexity in local atomic structures caused by random distribution of elements in a multinary alloy [38], CCAs with a single solid solution phase and the atoms on the sites of a specific Bravais lattice (e.g., face-centered cubic or body-centered cubic) exhibit many attractive functional properties [31,33,39]. For example, the high electrical conductivity and low thermal conductivity in CCAs such as Al$_x$CoCrFeNi and Pb-Sn-Te-Se, makes them promising for thermoelectric applications [40-43]. Furthermore, CCAs such as FeCoNi(AlSi)$_x$ are also promising candidates as soft magnetic materials due to their high saturation magnetization, high electrical resistivity, and high malleability [33]. The presence of several different exchange interactions in CCAs causes sluggish magnetic phase transitions and enhances the magnetocaloric effect (MCE) of these materials [44,45]. In addition, due to their multi-component nature, CCAs such as FeCoNiCuMn of certain stoichiometries have been shown to have a tunable Curie temperature $T_C$ that reaches room temperature [45], and the combination of tunable $T_C$ and enhanced MCE makes these CCAs attractive as magnetic refrigerant materials [44,45].

More recently, research has emerged focusing on using 2D CCAs such as $Mo_{1-x}W_xS_2$ and $(Al_xGa_{1-x})_{0.5}In_{0.5}P$ in energy conversion applications, such as photovoltaics, photocatalysts, and optoelectronic devices [9,10,46]. The entropic effect benefits the design and fabrication of multiple-component 2D CCAs with similar concentrations of the constituent elements. Instead of conventional 2D materials with fixed structural and electronic properties, this multi-component design opens up opportunities for tunable properties such as lattice constants and band gaps [47,48]. Desired phases and stoichiometries can be achieved through adjusting the contents of each element to enhance materials properties such as catalytic activity and light conversion efficiency [49]. For example, in order to acquire an optimal range of band gaps and thus optimized absorption coefficients, ternary [50,51], quaternary [52-54], and even penternary [55,56] CCAs have been computationally simulated and fabricated for photovoltaic applications. Moreover, the alloying method is able to modify the lattice constant of CCAs based on their constituent elements. For example, lattice strain can occur in van der Waals heterostructures due to lattice mismatch. The 2D CCAs from alloying with the components of similar crystal structures and in-plane lattice constants can bring in heterostructures and multijunction of the materials with small lattice mismatch, which lowers the interface strain in a heterostructure [57-59].

Quaternary 2D TMC alloys present themselves as potential CCAs due to their multi-elemental composition and high configurational entropy induced by alloying from binary TMCs. Because of tunable structural and electronic properties of multinary TMCs, it is important to investigate the thermodynamically favorable 2D compositionally complex transitional metal chalcogenide alloys (CCTMCAs) for high efficiency energy conversion applications such as photovoltaics, photocatalysts, and optoelectronics, as well as the spintronic applications of random-access memory (RAM) [11,12,60-64]. For example, quaternary 2D CCTMCAs such as $Cd_{1-x}Zn_xO_yS_{1-y}$ are found suitable in photovoltaics applications because of their high carrier mobilities and suitable range of direct band gaps [65-67]. In photovoltaics, the efficiency of a device is linked to effective carrier masses, which relate to the charger extraction and recombination dynamics and control the open-circuit voltage [13,68,69]. The carrier mobility, as one of the key properties of photovoltaics, depends on both the momentum relaxation time and effective mass, where the momentum relaxation time is inversely linked to the effective mass in lattice scattering [70]. It is shown that a large effective mass results in a decreased charge carrier mobility, which therefore lowers the efficiency of light conversion in photovoltaics [71]. Moreover, quaternary 2D CCTMCAs such as $Cu_2Mo(S_ySe_{1-y})_4$ are proposed as potential photocatalysts [72]. In the process of electrochemical water splitting, the band gap of a 2D CCTMCA determines the acceptable photon frequency during light adsorption, whereas the band alignments of conduction band minimum (CBM) and valence band maximum (VBM) are also considered essential in matching potentials of hydrogen/oxygen evolution reactions ($H_+/H_2$, $H_2O/O_2$) energy at different pH levels [62,73]. Additionally, 2D CCTMCAs are also found as potential spintronic applications in spin-logic devices such as RAM for their strong SOC. These functions rely on the controlling of the electron/hole spins, which comes from the metal *d*-orbital states in the heavy metal atoms in 2D CCTMCAs [64,74,75]. To sum up, 2D CCTMCAs play significant roles in the applications of energy conversion and spintronics. In order to understand the dependency of various properties of 2D CCTMCAs and their effects in those applications, it is essential to characterize the properties of 2D CCTMCAs such as lattice constants, band alignment, effective carrier masses, spin orbit splitting, and so on.

Different methods such as Korringa–Kohn–Rostoker coherent-potential-approximation (KKR-CPA) method and density functional theory (DFT) calculation have been used to study various properties such as lattice parameters, band gap, and band alignment in CCAs [9,16,76]. Reports using the KKR-CPA method to study bulk CCAs have confirmed a reduction in electron mean free path and subsequent decrease in electrical and thermal (from electronic contributions) conductivities with increasing principal elements [37]. Similarly, the KKR-CPA method has been used to study the electronic, magnetic, and transport properties of the Fe-intercalated bulk $TaS_2$ TMC alloys [77]. DFT calculations, on the other hand, have been widely used in studying the properties of 2D CCTMCAs such as band gaps and phase stability. During DFT calculation, models of random alloy and special quasi-random structure (SQS) are proposed in order to simulate the disorder CCAs. For example, the DFT calculations using a random alloy model have shown tunable band gaps of quaternary $Mo_{1-x}W_xS_{2y}Se_{2(1-y)}$ 2D CCTMCAs dependent on the composition, which are consistent with the experiment [9]. DFT calculations on the same 2D CCTMCAs have also demonstrated a spinodal decomposition from miscibility gap and the formation of lateral heterostructures at certain compositions of the alloy. This decomposition is also confirmed from experimentally observed phase segregation due to the miscibility gap [16]. Moreover, the combination of DFT using SQS models to study the formation enthalpies of ternary TMC alloys has shown that $MoSe_{2(1-x)}Te_{2x}$, $WSe_{2(1-x)}Te_{2x}$, $MoS_{2(1-x)}Te_{2x}$, and $WS_{2(1-x)}Te_{2x}$ alloy systems are unstable at 0 K [76]. The SQS approach has also been used to form disordered alloy models for ternary 2D CCTMCAs alloys, where using the SQS approach to calculate bowing parameters has shown that the in-plane lattice constant varies almost linearly with changes in the composition of ternary TMC alloys [76]. Comparing to the KKR-CPA method, DFT calculations can predict more accurate structural and electronic properties such as lattice constants, electronic density of states, and band gaps [78-80]. However, when dealing with quaternary 2D CCTMCAs with different compositions, supercells with a number of atoms are required, which makes the DFT calculations very time-consuming. Therefore, alternative methods are desired in efficiently characterizing structural and electronic properties.

Virtual crystal approximation (VCA), as an alternative method to investigate the CCAs of different compositions, has been applied to reduce computational cost while achieving a comparative accuracy to supercell-DFT calculations using an averaged potential from mixing elemental potentials [81]. Specifically, the VCA method provides a convenient and efficient way to model CCAs by generalizing their multi-elemental composition into the weighted average of the individual alloying elements [82]. In comparing to regular DFT calculations of large supercells in investigating multinary CCAs, DFT calculation using VCA method (VCA-DFT) neglects any short-range order and local distortions, and therefore cannot replicate the fine details of an alloy, this method offers much greater simplicity and lower computational cost. The VCA-DFT method as implemented in the Vienna Ab initio Simulation Package (VASP) has found successful applications in studying TMC systems such as $WSe_{2(1-x)}Te_{2x}$ [83]. The VCA-DFT method has also been used to obtain structural properties [84,85], phase determination [86], and electronic properties such as band gaps and effective masses of carriers [87,88], the results of which are all comparable to experiments [89,90]. Therefore, utilizing the VCA-DFT method is helpful in characterizing the CCAs of different compositions, facilitating materials screening and selection for different applications.

In this work, we propose a high-throughput workflow to investigate the properties of 2D CCTMCAs as candidates for various energy and information technology applications such as light conversion and computer logic systems. We use the Mo-W-S-Se, Mo-W-S-Te, and Mo-W-Se-Te 2D CCTMCAs as examples to illustrate the search and selection of 2D CCTMCAs with different structural and electrical properties via VCA-DFT calculations. Meanwhile, we adopt the unit cell and SQS models to benchmark the accuracy of the results from using the VCA-DFT method. We show that the structural and electronic properties of 2D CCTMCAs including in-plane lattice constants, band gaps, effective carrier masses, spin-orbital splitting, and band alignment are in good agreement with the DFT calculations based on the unit-cell (unit-cell-DFT) and SQS (SQS-DFT) models. We then select nine ternary and three quaternary 2D CCTMCAs to investigate the stability using the metrics of formation energy and Gibbs free energy. We then focus on the three quaternary 2D CCTMCAs, $Mo_{0.5}W_{0.5}SSe$, $Mo_{0.5}W_{0.5}STe$, and $Mo_{0.5}W_{0.5}SeTe$, which are found to be able to form type II band alignments with other quaternary 2D CCTMCAs, owing to their high configuration entropy. These three quaternary 2D CCTMCAs have negative Gibbs free energies at 300 K, serving as good examples as entropy-stabilized multinary 2D alloys. Additionally, $Mo_{0.5}W_{0.5}SSe$, $Mo_{0.5}W_{0.5}STe$, and $Mo_{0.5}W_{0.5}SeTe$ show high electrical conductivity at room temperature (300K), making them possible to be utilized in energy conversion applications. We therefore suggest the VCA-DFT based workflow as a potential high-throughput framework in searching for 2D CCTMCAs for various applications.

**Methods**

We apply the VCA-DFT method in the Vienna Ab Initio Package (VASP) [91] to study the Mo-W-S-Se, Mo-W-S-Te, and Mo-W-Se-Te 2D CCTMCAs, $Mo_yW_{1-y}S_{2x}Se_{2(1-x)}$, $Mo_yW_{1-y}S_{2x}Te_{2(1-x)}$, and $Mo_yW_{1-y}Se_{2x}Te_{2(1-x)}$, with $x$ and $y$ ranging from 0 to 1 at an incremental step of 0.05. The range of *x* and *y* results in three sets of 441 quaternary 2D CCTMCAs. Figure 1(a) illustrates the structure model from the VCA method, and Figure 1(b) and (c) display a $4 \times 4 \times 1$ SQS supercell obtained from the ATAT package [92]. We perform all the calculations using VASP, and the plane waves of kinetic energies smaller than 500 eV are included in the basis sets. We use the standard projector augmented wave (PAW) potential files [93,94] for Mo, W, S, Se, and Te. The Perdew-Burke-Ernzerhof (PBE) functional [95] in DFT is well known to underestimate band gaps [96,97]. If more advanced theories accounting for many-body effects that are missing in the PBE functional are used, the theoretical band gaps will generally be higher. However, the underestimated band gaps will be compensated for by the absence of SOC in the calculations. The net effect is that, for example, the PBE band gaps of WSSe and MoSSe agree well with the experimental band gaps [98]. We use a Monkhorst-Pack [99] $15 \times 15 \times 1$ *k*-point grid. The in-plane lattice constants and atomic coordinates of all systems are fully optimized using a quasi-Newton algorithm with the force convergence criterion of 0.01 eV/Å. We calculate the band structures along the $\Gamma \rightarrow M \rightarrow K \rightarrow \Gamma$ special *k*-point path (each line segment has 40 *k* points), where we extract the band gaps and effective carrier masses. We also calculate effective electron and hole masses using the Sumo package [100] and report the averaged effective masses in the *K-M* and *K-Γ* directions. To calculate the spin-orbit splitting magnitude, we include SOC in the calculations. We calculate the electrical conductivities of three quaternary 2D CCTMCAs using the BoltzTraP module as implemented in the Pymatgen library [101,102].

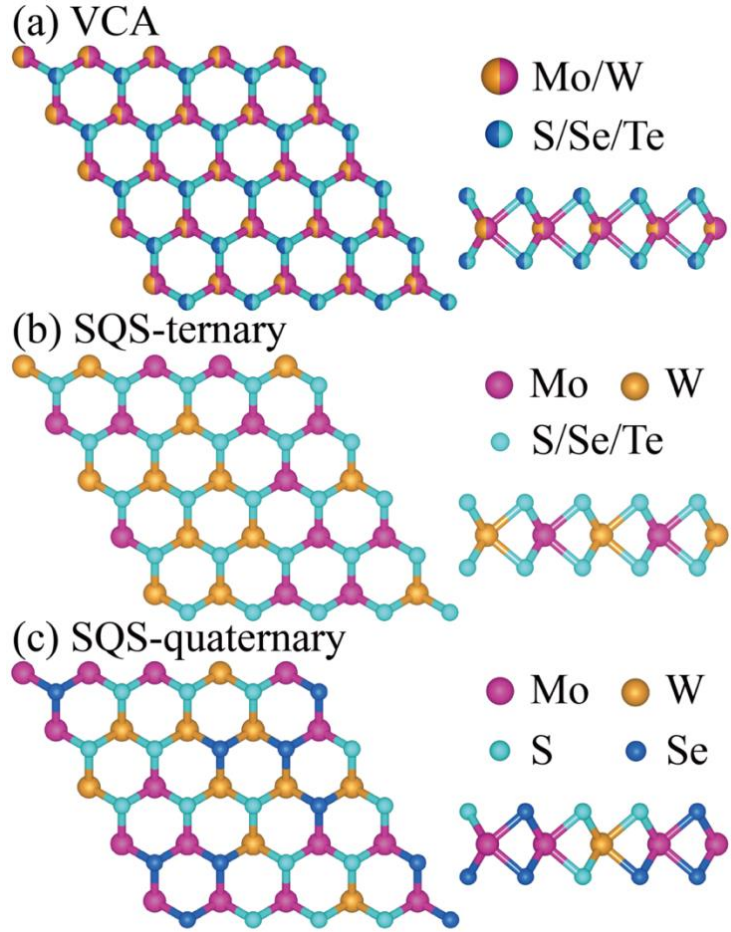

**Figure 1.** Top and side views of the atomic structure of 2D CCTMCAs based on (a) the VCA method, and (b) and (c) the SQS method for ternary and quaternary 2D CCTMCAs, respectively.

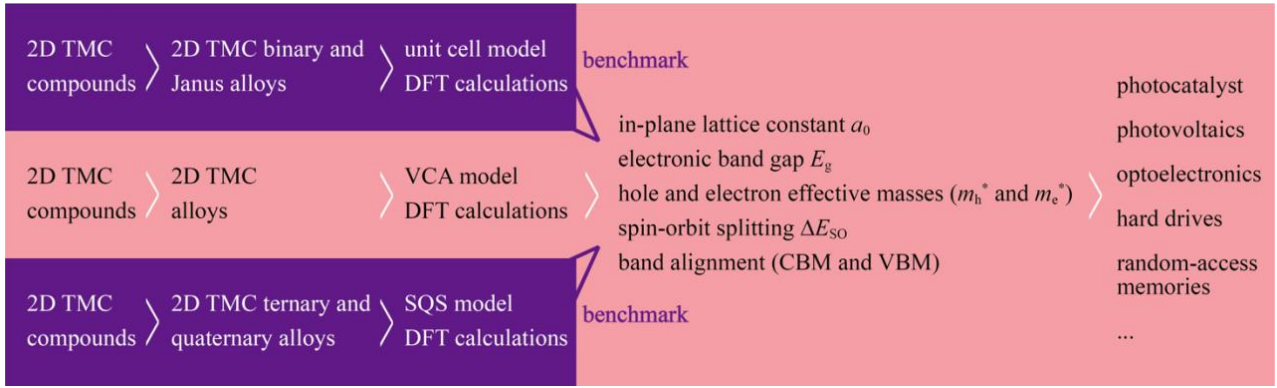

**Figure 2.** High-throughput workflow for designing 2D CCTMCAs.

Many previous computational studies have shown that 2D TMCs have potential in energy conversion and spintronic applications due to their excellent structural and electronic properties [11,12,60-64]. However, a systematic workflow that can be applied to search for materials for various applications in an efficient manner is lacking. In order to efficiently discover those materials from 2D TMCs, we propose a workflow (see Figure 2) that utilizes VCA-DFT to obtain essential structural and electronic properties. The workflow consists of firstly proposing several simple binary TMC compounds. Different alloying 2D CCTMCAs can thus be generated by the combination of the binary compounds. Then the workflow applies DFT calculations using the averaged pseudopotential of corresponding elements to characterize the VCA model. We compute five essential properties for selecting materials in various applications, which are in-plane lattice constant, band gap, hole and electron effective masses, spin-orbital splitting, and band alignment of CBM and VBM. At the same time, the reliability and accuracy of the calculation process is cross examined by benchmarking DFT calculations using unit-cell and SQS models. We use three-atom unit cells to model the binary $MX_2$ and Janus $MXY$ ($X \neq Y =$ S, Se, or Te) structures of TMCs, whereas for the complex ternary and quaternary 2D CCTMCAs, we create SQS supercells to simulate the disordered structures. Specifically, we validate our VCA-DFT calculations using six binary TMC unit cells and six Janus unit cells, as well as three ternary SQS supercells and three quaternary SQS supercells with special stoichiometries that are $Mo_{0.5}W_{0.5}S_2$, $Mo_{0.5}W_{0.5}Se_2$, $Mo_{0.5}W_{0.5}Te_2$, $Mo_{0.5}W_{0.5}SSe$, $Mo_{0.5}W_{0.5}STe$, and $Mo_{0.5}W_{0.5}SeTe$.

**Results and Discussion**

We perform VCA-DFT calculations to obtain five essential structural and electronic properties including the in-plane lattice constant, band gap, hole and electron effective masses, spin-orbit splitting, and band alignment, in order to search for suitable 2D CCTMCAs in potential applications of energy conversion and spin-logic devices. First of all, the in-plane lattice constant of a 2D material is a fundamental parameter in describing the monolayer geometry. Besides, the lattice constants of 2D CCTMCAs are helpful to understand the composition-dependent lattice change, which is important in considering the lattice matching of two different 2D CCTMCAs during designing the 2D stacked heterostructure, in order to avoid the misfit and change of the crystal structure. Secondly, it is essential to tune band gaps to fit in the range of photonic frequency of different light sources in light adsorption devices in photovoltaics and photocatalysts. An appropriate band gap range matching photonic frequency increases the efficiency of harvesting light energy. Thirdly, effective masses of electrons and holes are critical in predicting carrier optical response and transport property of semiconductors [103,104]. The effective masses also determines the effective density of states, which further impacts the open circuit voltage [71]. Fourthly, 2D Group-VI TMCs, such as $MoS_2$, $MoSe_2$, $WS_2$, and $WSe_2$, have been heavily investigated as potential materials in the field of spintronics and valleytronics due to their broken inversion symmetry [105-108]. These 2D materials contain two inequivalent valleys that occur at the $+K$ and $–K$ points at the edges of the first Brillouin zone, and time reversal symmetry in TMCs causes spin splitting with opposite spin signs at the $+K$ and $–K$ valleys, resulting from strong SOC in TMCs [105]. Due the broken inversion symmetry in TMCs, this coupling between spin and valley pseudospin causes the splitting of valence bands [105], and in $MoS_2$ specifically, SOC interactions have been shown to split the valence bands by around 0.16 eV [106,107]. The strong SOC in TMCs has been shown to allow for higher spin and valley polarization lifetime along with the manipulation of spin through valley properties [105]. Finally, band alignment describes an electronic property about a material's light conversion efficiency. In the design of thin film photovoltaics, the adjustment of conduction band alignment to a desired range is regarded as one of the most important factors to reach a high conversion efficiency [109,110]. Specifically, in combining two materials for photovoltaics, a heterojunction between *p*-type and *n*-type semiconductors needs to be utilized to provide a type II band alignment to provide pathways for exciton diffusion, separation, and dissociation, and carrier transportation [111-114].

Following the above high-throughput procedure, we begin with computing the five properties of $MX_2$ ($M$ = Mo, W; $X$ = S, Se, or Te) using unit-cell-DFT calculation as a benchmark to the corresponding VCA-DFT results. All 2D $MX_2$ in this work are assumed to adopt the 2$H$ phase. Table 1 lists these calculated properties for the six $MX_2$. We can see that the VCA-DFT method leads to identical results compared to the DFT results using a 3-atom unit cell. We also include the results from the literature for comparison. The VCA-DFT values for the five properties also agree well with previous studies. For example, the in-plane lattice constants and band gaps of MoS$_2$ (WSe$_2$) 3.18 Å (3.32 Å) and 1.67 eV (1.55 eV) from our VCA-DFT calculations, are nearly identical to 3.18 Å (3.32 Å) and 1.68 eV (1.53 eV) in the Refs. [115,116]. We also notice that the in-plane lattice constants of the six binary TMCs share three values, 3.18, 3.32, and 3.55 Å, where the binary TMCs of same chalcogen element have the same lattice constants. This is understood by the large difference in the anion radii of S$_2$- (1.70 Å), Se$_2$- (1.84 Å), and Te$_2$- (2.07 Å), whereas the cations of Mo$_{4+}$ and W$_{4+}$ possess similar ionic radii of 0.79 and 0.80 Å, respectively [117]. Moreover, the discrepancy in the covalent radii between Mo (1.54 Å) and W (1.62 Å) is also smaller than that among S (1.05 Å), Se (1.20 Å), and Te (1.38 Å) [118]. For the CBM and VBM, the VCA-DFT results match well with other DFT calculation results at the PBE level. Additionally, the CBM values of MoTe$_2$, WS$_2$, and WSe$_2$ are closer to the DFT calculation results in Ref. [119] using the PBE functional including SOC.

**Table 1.** In-plane lattice constant $a_0$, band gap $E_g$, hole and electron effective masses $m_h^*$ and $m_e^*$, spin-orbit splitting $\Delta E_{SO}$, conduction band minimum (CBM) and valencene band maximum (VBM) with reference to the vacuum level

of $MX_2$ with the 2$H$ structure. The first row of each property is obtained from using the VCA-DFT method, whereas the second row is from the literature.

|  | $MoS_2$ | $MoSe_2$ | $MoTe_2$ | $WS_2$ | $WSe_2$ | $WTe_2$ |
|---|---|---|---|---|---|---|
| $a_0$ (Å) | 3.18 | 3.32 | 3.55 | 3.18 | 3.32 | 3.55 |
|  | 3.18[a] | 3.32[b] | 3.55[c] | 3.18[b] | 3.32[b] | 3.55[d] |
| $E_g$ (eV) | 1.67 | 1.44 | 1.08 | 1.81 | 1.54 | 1.07 |
|  | 1.68[b] | 1.43[b] | 1.06[b] | 1.81[b] | 1.53[b] | 1.08[e] |
| $m_h^*$ ($m_0$) | 0.59 | 0.66 | 0.71 | 0.43 | 0.46 | 0.44 |
|  | 0.58[f] | 0.67[f] | 0.70[f] | 0.42[g] | 0.51[g] | 0.42[h] |
| $m_e^*$ ($m_0$) | 0.49 | 0.56 | 0.58 | 0.32 | 0.35 | 0.33 |
|  | 0.49[f] | 0.56[f] | 0.57[f] | 0.31[g] | 0.39[g] | 0.33[h] |
| $\Delta E_{SO}$ | 0.15 | 0.19 | 0.22 | 0.43 | 0.47 | 0.48 |
|  | 0.15[i] | 0.19[i] | 0.22[i] | 0.43[i] | 0.47[i] | 0.49[i] |
| CBM | -4.27 | -3.89 | -3.80 | -3.89 | -.3.56 | -3.63 |
|  | -4.29[j] | -3.89[j] | -3.74[j] | -4.09[j] | -3.69[j] | -3.61[j] |
| VBM | -5.94 | -5.33 | -4.89 | -5.70 | -5.11 | -4.70 |
|  | -5.98[j] | -5.31[j] | -4.84[j] | -5.87[j] | -5.20[j] | -4.72[j] |

[a]Ref. [115]
[b]Ref. [116]
[c]Ref. [120]
[d]Ref. [53]
[e]Ref. [121]
[f]Ref. [122]
[g]Ref. [123]
[h]Ref. [124]
[i]Ref. [125]
[j]Ref. [126]

**Table 2.** In-plane lattice constant $a_0$, band gap $E_g$, hole and electron effective masses $m_h^*$ and $m_e^*$, spin-orbit splitting $\Delta E_{SO}$, conduction band minimum (CBM) and valencene band maximum (VBM) with reference to the vacuum level of Janus $MXY$ with the $2H$ structure. The first row of each property is obtained from using the VCA-DFT method, whereas the second row is from using three-atom unit cells to simulate monolayer Janus structures.

|  | MoSSe | WSSe | MoSTe | WSTe | MoSeTe | WSeTe |
|---|---|---|---|---|---|---|
| $a_0$ (Å) | 3.26 | 3.26 | 3.40 | 3.39 | 3.45 | 3.47 |
|  | 3.25 | 3.25 | 3.36 | 3.36 | 3.43 | 3.43 |
| $E_g$ (eV) | 1.54 | 1.66 | 1.31 | 1.40 | 1.23 | 1.29 |
|  | 1.56 | 1.69 | 1.03 | 1.24 | 1.27 | 1.35 |
| $m_h^*$ ($m_0$) | 0.64 | 0.45 | 0.73 | 0.49 | 0.73 | 0.49 |
|  | 0.68 | 0.48 | 5.51 | 3.67 | 0.84 | 0.55 |
| $m_e^*$ ($m_0$) | 0.54 | 0.35 | 0.62 | 0.38 | 0.61 | 0.37 |
|  | 0.54 | 0.36 | 0.71 | 0.47 | 0.64 | 0.39 |
| $\Delta E_{SO}$ | 0.17 | 0.45 | 0.21 | 0.49 | 0.21 | 0.49 |
|  | 0.17 | 0.44 | 0.19 | 0.42 | 0.20 | 0.46 |
| CBM | -3.94 | -3.59 | -3.79 | -3.49 | -3.86 | -3.59 |
|  | -4.07 | -3.70 | -4.04 | -3.70 | -3.84 | -3.57 |
| VBM | -5.48 | -5.25 | -5.10 | -4.88 | -5.09 | -4.88 |
|  | -5.63 | -5.40 | -5.07 | -4.94 | -5.11 | -4.91 |

Many previous studies suggested that the VCA-DFT calculation is applicable to disordered semiconductor alloys [127,128], but this method does not consider the effects of lattice relaxation and assumes that the atoms are fixed at the ideal lattice sites [129]. We further benchmark VCA-DFT calculation by using the DFT calculation from unit cell models for Janus structure and SQS supercells for ternary structures (see Table 2 and Table 3) and quaternary systems (see Table 4) to ensure that the VCA-DFT results are consistent. Based on our results of VCA-DFT versus unit-cell-DFT and SQS-DFT calculations, the VCA-DFT method leads to comparable results such as lattice constants, band gaps, spin-orbital splitting, and band alignments with the unit-cell-DFT and SQS-DFT results. Some exceptions happen during validating the band gap and effective hole mass in the Janus MoSTe and WSTe structures, where the VCA-DFT calculations lead to the inconsistent conduction band as CBM, effective hole mass, and band gap compared to the DFT calculation using the unit cell models. Because its use of the averaged potential, the VCA-DFT is not always capable of predicting the local atomic environment [86,130,131], where a large difference in ionicity could result in low accuracy in predicting the electronic properties such as band alignment and band gap [131]. Moreover, when using the VCA-DFT in ground state calculation, the prediction of formation energy of 2D CCTMCAs shows the discrepancy from the SQS-DFT calculated result. Therefore, it is essential to use the unit-cell and SQS models to benchmark the VCA-DFT results. This not only validates the accuracy of VCA-DFT calculations on 2D CCTMCAs but also provides an indication when there is deviation in VCA-DFT results.

**Table 3**. In-plane lattice constant $a_0$, band gap $E_g$, hole and electron effective masses $m_h^*$ and $m_e^*$, spin-orbit splitting $\Delta E_{SO}$, conduction band minimum (CBM) and valencene band maximum (VBM) with reference to the vacuum level of ternary 2D CCTMCAs with the $2H$ structure. The first row of each property is obtained from using the VCA-DFT method, whereas the second row is from using 48-atom SQS supercells of monolayer ternary structures.

|  | $Mo_{0.5}W_{0.5}S_2$ | $Mo_{0.5}W_{0.5}Se_2$ | $Mo_{0.5}W_{0.5}Te_2$ |
|---|---|---|---|
| $a_0$ (Å) | 3.18 | 3.32 | 3.55 |
|  | 3.18 | 3.32 | 3.55 |
| $E_g$ (eV) | 1.74 | 1.50 | 1.09 |

|  | 1.71 | 1.46 | 1.06 |
|---|---|---|---|
| $m_h^*$ ($m_0$) | 0.51 | 0.56 | 0.57 |
|  | 0.49 | 0.51 | 0.55 |
| $m_e^*$ ($m_0$) | 0.40 | 0.46 | 0.44 |
|  | 0.33 | 0.38 | 0.44 |
| $\Delta E_{SO}$ | 0.29 | 0.33 | 0.35 |
|  | 0.29 | 0.30 | 0.24 |
| CBM | -4.11 | -3.73 | -3.72 |
|  | -4.12 | -3.76 | -3.73 |
| VBM | -5.84 | -5.23 | -4.81 |
|  | -5.83 | -5.22 | -4.79 |

**Table 4**. In-plane lattice constant $a_0$, band gap $E_g$, hole and electron effective masses $m_h^*$ and $m_e^*$, spin-orbit splitting $\Delta E_{SO}$, conduction band minimum (CBM) and valencene band maximum (VBM) with reference to the vacuum level of quaternary 2D CCTMCAs with the 2*H* structure. The first row of each property is obtained from using the VCA-DFT method, whereas the second row is from using 48-atom SQS supercells of monolayer quaternary structures.

|  | Mo$_{0.5}$W$_{0.5}$SSe | Mo$_{0.5}$W$_{0.5}$STe | Mo$_{0.5}$W$_{0.5}$SeTe |
|---|---|---|---|
| $a_0$ (Å) | 3.26 | 3.40 | 3.45 |
|  | 3.25 | 3.35 | 3.43 |
| $E_g$ (eV) | 1.60 | 1.35 | 1.26 |
|  | 1.57 | 1.29 | 1.23 |
| $m_h^*$ ($m_0$) | 0.54 | 0.61 | 0.61 |
|  | 0.52 | 0.53 | 0.55 |
|  | 0.44 | 0.49 | 0.48 |

| $m_e^*$ ($m_0$) | 0.49 | 0.59 | 0.37 |
| --- | --- | --- | --- |
| $\Delta E_{SO}$ | 0.31 | 0.35 | 0.36 |
| | 0.29 | 0.20 | 0.25 |
| CBM (eV) | -5.37 | -5.00 | -4.99 |
| | -5.48 | -5.04 | -4.93 |
| VBM (eV) | -3.76 | -3.64 | -3.72 |
| | -3.90 | -3.75 | -3.69 |

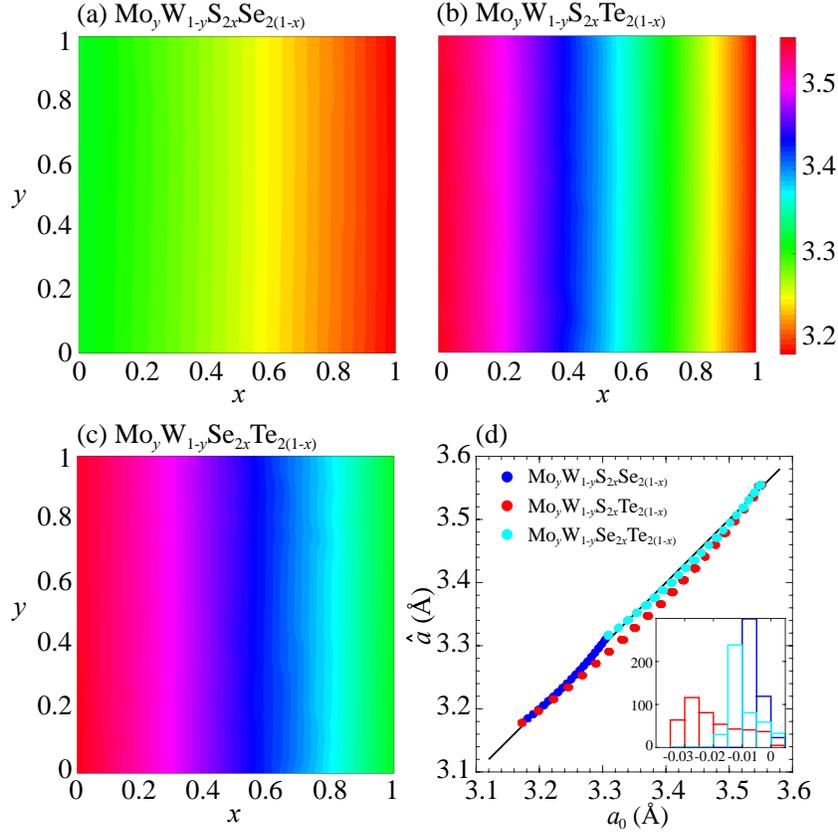

**Figure 3**. In-plane lattice constants of (a) Mo$_y$W$_{1-y}$S$_{2x}$Se$_{2(1-x)}$, (b) Mo$_y$W$_{1-y}$S$_{2x}$Te$_{2(1-x)}$, and (c) Mo$_y$W$_{1-y}$Se$_{2x}$Te$_{2(1-x)}$ 2D CCTMCAs calculated with the PBE functional. (d) Comparison between the in-plane lattice constants from the VCA-DFT method and from the estimation in Eq.2.

Figure 3(a)-(c) displays the in-plane lattice constants of 2D CCTMCAs Mo$_y$W$_{1-y}$S$_{2x}$Se$_{2(1-x)}$, Mo$_y$W$_{1-y}$S$_{2x}$Te$_{2(1-x)}$, and Mo$_y$W$_{1-y}$Se$_{2x}$Te$_{2(1-x)}$, and the range of in-plane lattice constants for each set of 2D CCTMCAs is summarized in Table 5. Consistent with the trend shown in Table 5, the in-plane lattice constants strongly depend on the content ($x$) of chalcogen, while the change in the content of transitional metal ($y$) almost has no effect. To further investigate the relationship between the composition and the lattice constant, we hypothesize the following chemical reaction:

$$\text{Mo}_y\text{W}_{1-y}\text{S}_{2x}\text{Se}_{2(1-x)} \rightarrow xy\text{MoS}_2 + (1-x)y\text{MoSe}_2 + x(1-y)\text{WS}_2 + (1-x)(1-y)\text{WSe}_2. \quad (1)$$

We the adopt Vegard's law [132] to link the 2D CCTMCAs with binary TMCs in Eq.1 and summarize the relationship between lattice constants $a$ of quaternary 2D CCTMCAs (e.g., Mo$_y$W$_{1-y}$S$_{2x}$Se$_{2(1-x)}$) and binary TMCs (e.g., MoS$_2$, MoSe$_2$, WS$_2$, or WSe$_2$). As an example, the lattice constant of Mo$_y$W$_{1-y}$S$_{2x}$Se$_{2(1-x)}$ can be written as

$$a = xya_{\text{MoS}_2} + (1-x)ya_{\text{MoSe}_2} + x(1-y)a_{\text{WS}_2} + (1-x)(1-y)a_{\text{WSe}_2} \quad (2)$$

that is, the lattice constants of quaternary 2D CCTMCAs can be approximated as the combination of compositionally dependent lattice constants of the binary TMCs. Figure 3(d) compares $a_0$ from VCA-DFT calculations and $a$ from Eq. 2. In the inset, we show the distribution of the deviation between two sets of data. The results from these two methods agree well with each other, with average deviations of 0.006, 0.020, and 0.010 Å for these three 2D CCTMCAs, respectively.

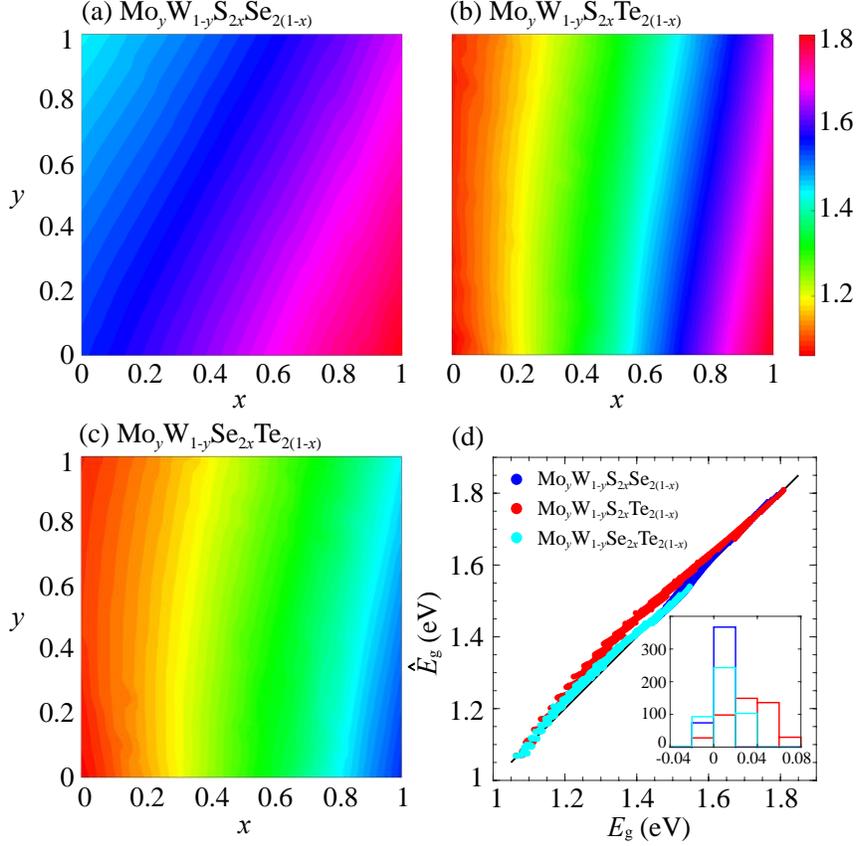

**Figure 4.** Band gaps of (a) Mo$_y$W$_{1-y}$S$_{2x}$Se$_{2(1-x)}$, (b) Mo$_y$W$_{1-y}$S$_{2x}$Te$_{2(1-x)}$, and (c) Mo$_y$W$_{1-y}$Se$_{2x}$Te$_{2(1-x)}$ 2D CCTMCAs calculated with the PBE functional. (d) Comparison between the band gaps from the VCA-DFT method and from the estimation in Eq.3.

**Figure 4**(a)-(c) display the band gaps of 2D CCTMCAs Mo-W-S-Se, Mo-W-S-Te, and Mo-W-Se-Te, which show that the band gaps range from 1.44 ~ 1.81 eV, 1.06 ~ 1.81 eV, and 1.06 ~ 1.55 eV, for the Mo-W-S-Se, Mo-W-S-Te, and Mo-W-Se-Te 2D CCTMCAs, respectively. We again apply Vegard's law [132] to estimate the band gaps of 2D CCTMCAs by using the band gaps of binary TMCs (see Table 1). Similar to Eq. 2, we can write the formula for the band gap using the binary TMCs. For example, for the Mo-W-S-Se 2D CCTMCAs, their band gaps can be written as,

$$E_g = xyE_{g,\text{MoS}_2} + (1-x)yE_{g,\text{MoSe}_2} + x(1-y)E_{g,\text{WS}_2} + (1-x)(1-y)E_{g,\text{WSe}_2} \qquad (3)$$

and the other two 2D CCTMCA systems have the similar formulas. Figure 4(d) compares the band gaps resulted from VCA-DFT calculations and from Eq.3. As can be seen, the calculated band gaps of the Mo-W-S-Se, Mo-W-S-Te, and Mo-W-Se-Te 2D CCTMCAs from the two methods are nearly identical, with the average deviations of merely 0.007, 0.039, and 0.011 eV respectively. This consistency indicates that using VCA-DFT can lead to reliable lattice constants and band gaps of 2D CCTMCAs in an efficient way.

The diverse lattice constants and band gaps of 2D CCTMCAs are essential for various applications such as in the design of heterostructures for light harvesting [50,133]. For example, the growth of heterostructures requires a lattice match between a 2D CCTMCA as the substrate layer and another 2D CCTMCA grown on the substrate. The matching in lattice constants of two monolayers of 2D CCTMCAs is beneficial for reducing the bilayer strain, whereas the wide band gap ranges of 2D CCMTCAs could also be helpful in maximizing the light conversion efficiency for many applications [76,134]. Figure 5 depicts the relationship between lattice constants and band gaps of the three 2D CCTMCAs. We know from the plot that there are majorly two regions based on heterostructure lattice matching, bounded by three edges of $MoS_2/WS_2$, $MoSe_2/WSe_2$, and $MoTe_2/WTe_2$ to the left, middle, and right edges. The two regions correspond to $Mo_yW_{1-y}S_{2x}Se_{2(1-x)}/Mo_yW_{1-y}S_{2x}Te_{2(1-x)}$, and $Mo_yW_{1-y}Se_{2x}Te_{2(1-x)}/Mo_yW_{1-y}S_{2x}Te_{2(1-x)}$ heterostructures, respectively. The lattice constant and band gap distributions of three 2D CCTMCAs in Figure 5 therefore provides guidance in designing heterostructures (see below) from a pair of 2D CCTMCA monolayers in order to minimize the lattice mismatch as well as to maintain the desired band offset value.

**Table 5**. Ranges of in-plane lattice constant $a_0$, band gap $E_g$, hole and electron effective masses $m_h^*$ and $m_e^*$, spin-orbit splitting $\Delta E_{SO}$, conduction band minimum (CBM) and valencene band maximum (VBM) with reference to the vacuum level of quaternary 2D CCTMCAs with the 2$H$ structure.

|  | $Mo_yW_{1-y}S_{2x}Se_{2(1-x)}$ | $Mo_yW_{1-y}S_{2x}Te_{2(1-x)}$ | $Mo_yW_{1-y}Se_{2x}Te_{2(1-x)}$ |
|---|---|---|---|
| $a_0$ (Å) | 3.18 ~ 3.32 | 3.18 ~ 3.55 | 3.32 ~ 3.55 |
| $E_g$ (eV) | 1.44 ~ 1.81 | 1.06 ~ 1.81 | 1.06 ~ 1.55 |
| $m_h^*$ ($m_0$) | 0.43 ~ 0.67 | 0.43 ~ 0.73 | 0.44 ~ 0.73 |
| $m_e^*$ ($m_0$) | 0.32 ~ 0.57 | 0.32 ~ 0.62 | 0.32 ~ 0.61 |
| $\Delta E_{SO}$ | 0.15 ~ 0.47 | 0.15 ~ 0.49 | 0.19 ~ 0.49 |
| CBM (eV) | -4.27 ~ -3.55 | -4.32 ~ -3.46 | -3.94 ~ -3.56 |
| VBM (eV) | -5.94 ~ -5.10 | -5.94 ~ -4.68 | -5.33 ~ -4.70 |

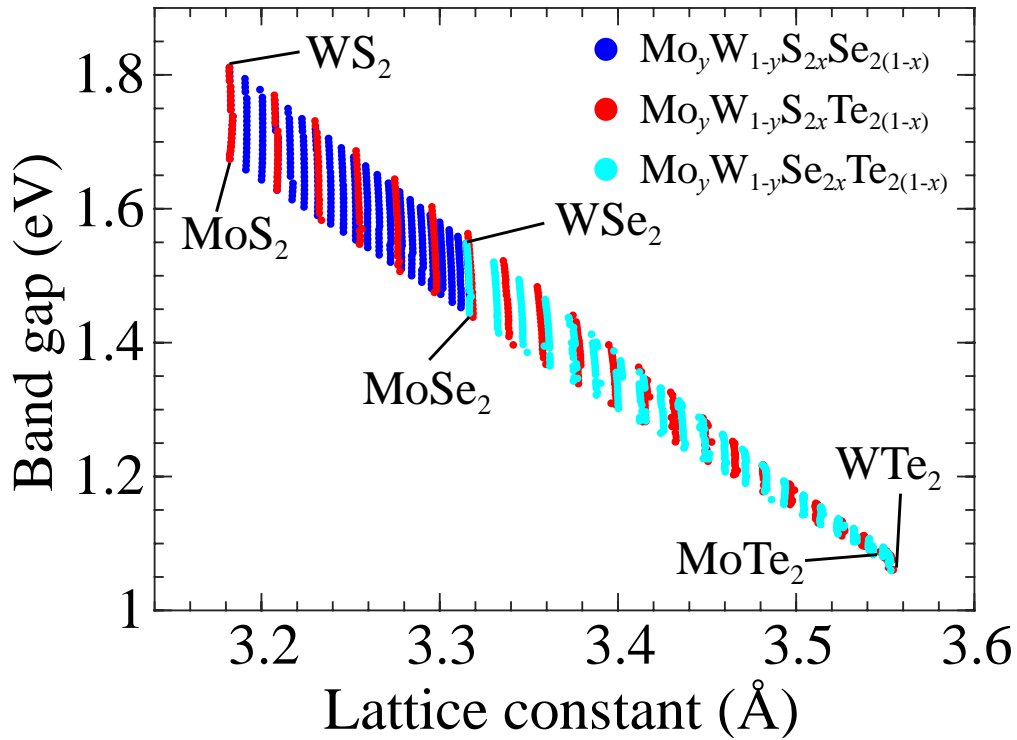

**Figure 5**. Relation between the predicted band gaps and the lattice constants of three sets of 2D CCTMCAs calculated with the PBE functional.

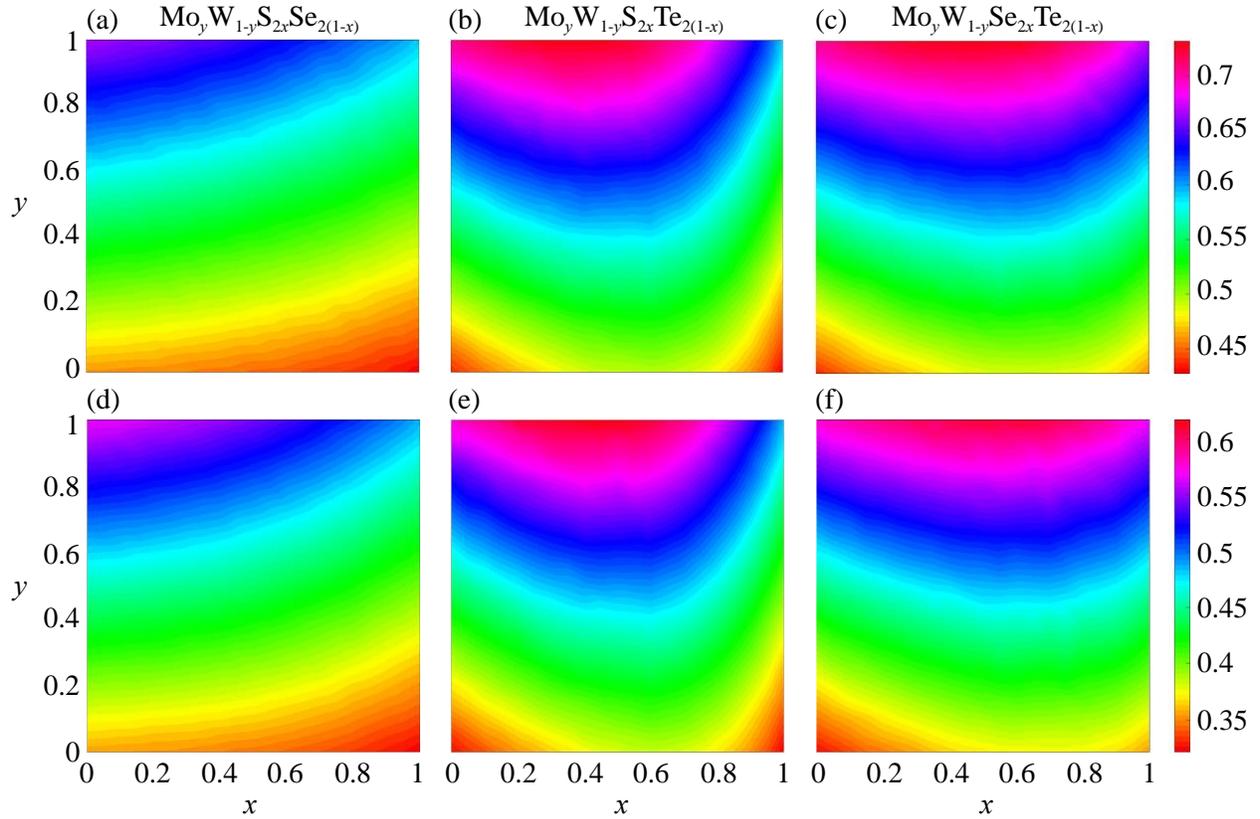

**Figure 6**. Hole effective masses of (a) $Mo_yW_{1-y}S_{2x}Se_{2(1-x)}$, (b) $Mo_yW_{1-y}S_{2x}Te_{2(1-x)}$, and (c) $Mo_yW_{1-y}Se_{2x}Te_{2(1-x)}$ 2D CCTMCAs, calculated with the PBE functional. The corresponding electron effective masses are shown in (d), (e), and (f).

Figure 6 displays the hole and electron effective masses of 2D CCTMCAs. The four corners of Figure 6(a) stand for the effective carrier masses of $MoS_2$, $MoSe_2$, $WS_2$, and $WSe_2$. We observe that increasing the W content (i.e., decreasing $y$ while $x$ is fixed) generally reduces the effective masses. For the Mo-W-S-Se 2D CCTMCAs, increasing the Se content lowers the effective masses of both electrons and holes. The relationship between the effective masses and two anion contents ($x$ and 1-$x$ value) is complicated in the other two 2D CCTMCAs. The lower effective carrier mass is found at the two regions of $x$ close to 0 and $x$ close to 1, where one of the chalcogens dominates in the 2D CCTMCAs. For example, in the Mo-W-S-Te 2D CCTMCAs, the effective carrier mass first increases as the content of S increases from 0 to 0.6, and then decreases as the content of S further increases up to 1.0. In designing heterostructures, low effective masses are beneficial to enhance the carrier transport and thus improve the collected photocurrent during the light harvesting process [71,135,136]. However, too small effective carrier masses are associated with a large curvature of electronic dispersion and thus sharp band edges, affecting the local density of states. As a result, the overall collected photocurrent will decrease, thus degrading the efficiency of light conversion [137]. Therefore, small effective carrier masses and local carrier concentration are competing with each other to maintain an optimal light conversion of photovoltaics and photocatalysts. The compositional variation of 2D CCTMCAs results in the tunable effective mass ranges for both electrons and holes. The design using compositionally complex systems offers a promising method in adjusting the effective carrier mass for efficient light adsorption devices.

Figure 7 shows the calculated spin-orbit splitting of 2D CCTMCAs. Compared to monolayer $MoS_2$, which has a $\Delta E_{SO}$ of 0.15 eV [125], the 2D CCTMCAs generally can have higher values of $\Delta E_{SO}$ and up to more than three times of the value of $MoS_2$. As a proposed candidate for spintronics devices, graphene shows a $\Delta E_{SO}$ of about 0.01 eV [138], which is much lower than those of the 2D CCTMCAs, suggesting that the 2D CCTMCAs has the potential in spintronic applications for spin-logic devices such as RAM [139]. Strong spin-orbit splitting and its insensitivity to anion species in 2D CCTMCAs results from the presence of out-of-plane mirror symmetry and absence of inversion symmetry [140,141]. As a consequence, the resulting electric field is generated in the plane of cations causing electrons to move in the same plane. The SOC interactions split the energy degeneracy of these electrons and the splitting magnitude depends on only the atomic numbers of the cation species (Mo and W, and the latter is much heavier, so is the stronger SOC and $\Delta E_{SO}$). Figure 7 also reveals that $\Delta E_{SO}$ is more dependent on the content of cations than that of anions. By increasing the content of the cations, e.g., reducing $y$, the $\Delta E_{SO}$ increases rapidly from 0.15 eV to nearly 0.50 eV. By contrast, increasing *x* does not alter *ΔE*SO as much as change *y*. The strong spin-orbit splittings of 2D CCTMCAs suggest the possibility of utilizing them for spintronic applications. This is understood from the source of SOC from the interaction between electron and magnetic field induced by the nucleus spin. Because the magnetic field is directly related to the charge from the nucleus, a larger atomic number will have a stronger SOC [140,142,143].

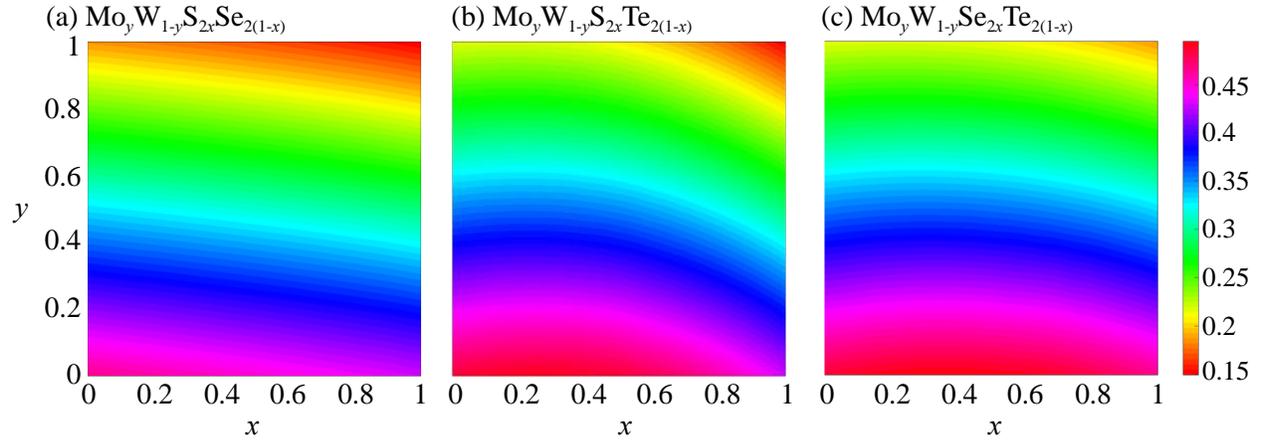

**Figure 7**. Spin-orbit splitting of (a) $Mo_yW_{1-y}S_{2x}Se_{2(1-x)}$, (b) $Mo_yW_{1-y}S_{2x}Te_{2(1-x)}$, and (c) $Mo_yW_{1-y}Se_{2x}Te_{2(1-x)}$ 2D CCTMCAs calculated with the PBE functional.

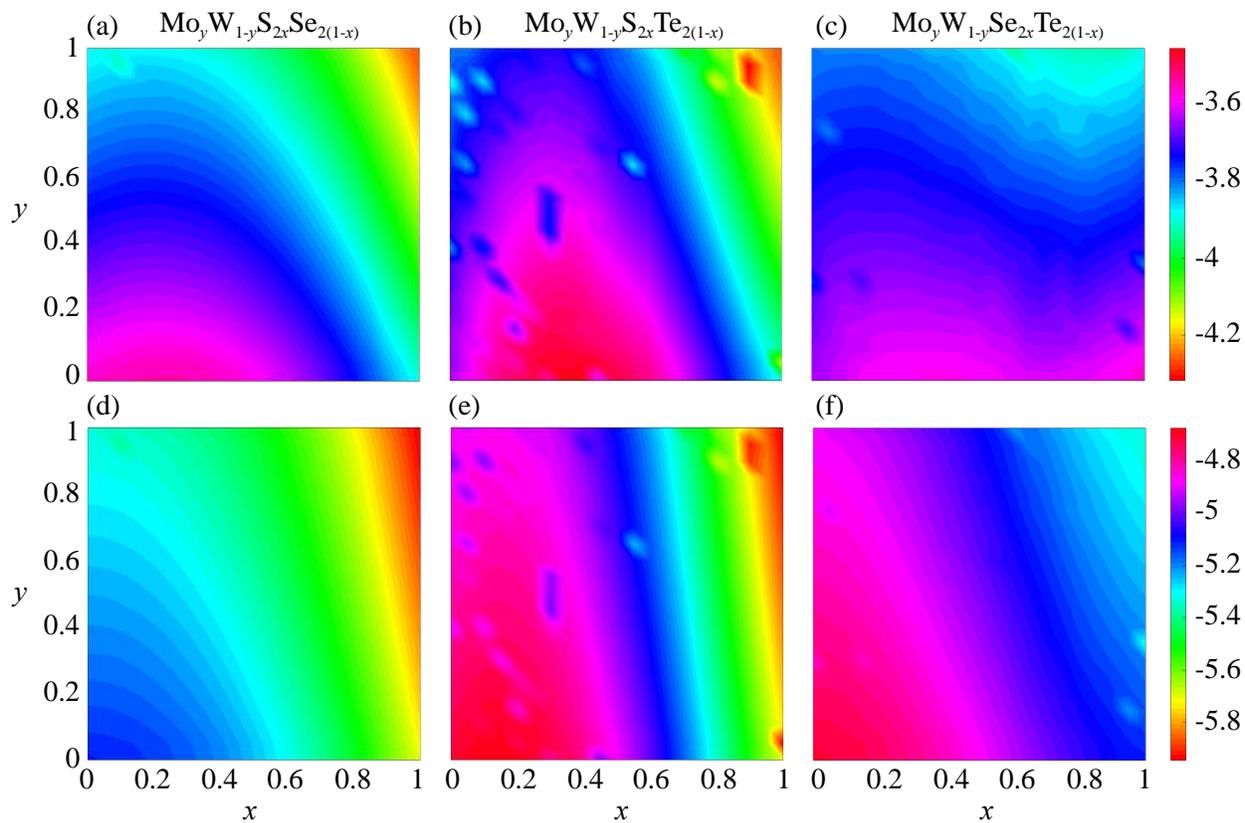

**Figure 8**. Conduction band minima of (a) $Mo_yW_{1-y}S_{2x}Se_{2(1-x)}$, (b) $Mo_yW_{1-y}S_{2x}Te_{2(1-x)}$, and (c) $Mo_yW_{1-y}Se_{2x}Te_{2(1-x)}$ 2D CCTMCAs. The corresponding valence band maxima are shown in (d), (e), and (f), calculated with the PBE functional.

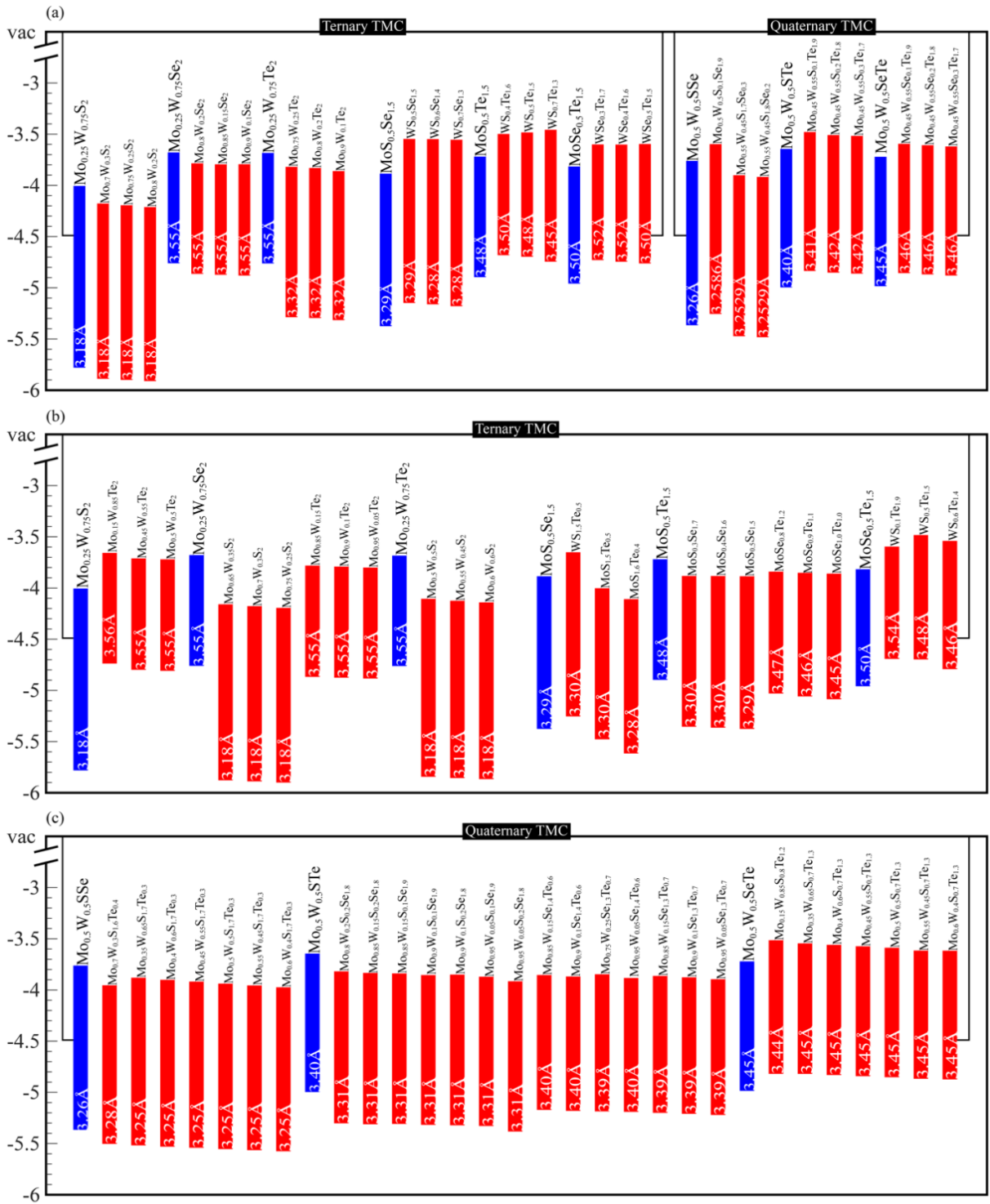

**Figure 9**. Band alignment of ternary and quaternary 2D CCTMCAs. The red bars are selected to compare to the reference blue bars with ((a) or (b)) the same elements and (c) different elements. The stoichiometry of each 2D

CCTMCA is shown above the bars and the number at the bottom of each bar denotes the corresponding in-plane lattice constant.

Figure 8 shows the CBM and VBM of 2D CCTMCAs with reference to the vacuum level. The conduction band offset (CBO) and valence band offset (VBO) are calculated, respectively, as the differences in the CBMs and VBMs of two 2D CCTMCAs. From the maximum and minimum values of CBM and VBM in the three sets of 2D CCTMCAs, we determine the ranges for CBO as from 0.049 to 0.861 eV and VBO as from 0.001 to 1.265 eV. Previous studies have shown that the stacking bilayer structure of TMCs can achieve ranges of CBO from $0.76 \pm 0.12$ eV and VBO from $0.83 \pm 0.07$ eV [144-147]. Because of the wide range of band offsets, it is possible to obtain different types of heterostructures using one 2D CCTMCA and pairing it with another 2D CCTMCA. Additionally, the band gap ranges between the CBOs and VBOs correspond to the photon frequencies within the near-infrared region of the solar spectrum, endowing the potential of employing 2D CCTMCAs for the applications of high-efficiency photovoltaics [148,149]. Figure 9(a) displays 27 selected CBM and VBM of 2D CCTMCAs of ternary or quaternary systems that can form type II heterostructures with the selected reference ternary and quaternary 2D CCTMCAs. For example, the CBO and VBO of the $Mo_{0.5}W_{0.5}STe/Mo_{0.95}W_{0.05}S_{0.2}Se_{1.8}$ heterostructure are 0.27 and of 0.39 eV, respectively.

At the same time, in the design of heterostructure 2D CCTMCAs, because of the composition-dependent wide range of CBO and VBO, versatile heterostructures are achievable by stacking different composition/type of 2D CCTMCAs together. The search of pairing 2D CCTMCAs to form stacked heterostructures should also account for the relationship between lattice constants and band gaps (see Figure 5). Specifically, in designing the heterostructure, to minimize the lattice mismatch, two systems with small difference in lattice constants are preferred. We here investigate two sets of 2D CCTMCAs of Mo-W-S-Se and Mo-W-S-Te, and Mo-W-S-Te and Mo-W-Se-Te, as shown in Figure 5, where the range of lattice constants of two 2D CCTMCAs in each set overlaps with each other. Figure 9 (b) and (c) shows the band alignments between some selected 2D CCTMCAs (in red color) that can match with the reference 2D CCTMCA (in blue color) in ternary and quaternary systems. From the results of both ternary and quaternary 2D CCTMCAs, we find many of them can form type II heterostructures with a small lattice mismatch from the reference material, indicating the potential of 2D CCTMCAs as building blocks for light harvesting heterostructures. Because the Mo-W-S-Se 2D CCTMCAs have a relatively small range of in-plane lattice constants (see Figure 3), Figure 9(b) shows that there is no ternary Mo/W-Se-Te 2D CCTMCA that satisfy both small lattice mismatch and type II band alignment with the reference of ternary Mo/W-S-Se 2D CCTMCA. The ternary Mo-W-S/Se and Mo-W-S/Te 2D CCTMCAs heterostructure shows that the largest lattice mismatch can reach to 10% between $Mo_yW_{(1-y)}S_2$ and $Mo_yW_{(1-y)}Te_2$ (e.g., the lattice mismatch in the $Mo_{0.25}W_{0.75}S_2/Mo_{0.15}W_{0.85}Te_2$ heterostructure is 11.9%). The ternary Mo/W-S-Te system has a relatively large range of lattice constants, making it easier to find the 2D CCTMCA monolayers from both Mo/W-S-Se and Mo/W-Se-Te that can show the type II alignment with each other with a negligible lattice mismatch. For example, the lattice mismatch is nearly zero for the $MoS_{0.5}Se_{1.5}/WS_{1.5}Te_{0.5}$ and $MoS_{0.5}Te_{1.5}/MoSe_{0.8}Te_{1.2}$ heterostructures. In quaternary 2D CCTMCAs (see Figure 9(d)), since there is more degree of freedom of 2D CCTMCA monolayers from both anion and cation content variations, the lattice mismatch between Mo-W-S-Se and Mo-W-S-Te 2D CCTMCAs drops to near 0%. There is still no Mo-W-S-Te and Mo-W-Se-Te 2D CCTMCA heterostructure that has a small lattice mismatch and can form a type II band alignment, which is consistent with Figure 5. Similar to ternary Mo/W-S-Te 2D CCTMCAs, the quaternary Mo-W-S-Te 2D CCTMCAs with a wide range of lattice constants can form the type II alignment of both Mo-W-S-Se and Mo-W-Se-Te 2D CCTMCA monolayers with almost zero lattice mismatch. The abundant choices in selecting 2D CCTMCA heterostructures with small lattice mismatch and suitable band offset range open up opportunities for a variety of heterostructure.

One concern in designing 2D CCTMCAs is their thermodynamic stability. Here we address this concern by focusing on the Gibbs free energy in nine ternary and three quaternary 2D CCTMCAs with equal $x$ and $y$. The three quaternary 2D CCTMCAs are modelled using the SQS method. We compute the Gibbs free energy, a combination of the ground-state energy of mixing at 0 K and the temperature-dependent energy from configurational entropy, to evaluate the stability of 2D CCTMCAs at different temperatures. We first compute the energy reference in the 2D CCTMCAs of quaternary 2D CCTMCAs such as $Mo_yW_{1-y}S_{2x}Se_{2(1-x)}$ by assuming the 2D CCTMCAs is made from four parent binary alloys of $MoS_2$, $MoSe_2$, $WS_2$, and $WSe_2$. Therefore, based on Vegard's law [132], the reference formation energy $E_{ref}$ is represented as a function of the ground state energies $E_f$ of the four binary TMCs and the coefficients from the content of each element, which can be written as,

$$E_{ref,1}(x,y) = xyE_{f,MoS_2} + y(1-x)E_{f,MoSe_2} + x(1-y)E_{f,WS_2} + (1-x)(1-y)E_{f,WSe_2} \quad (4)$$

Similarly, we have reference formation energies for the other two systems written as,

$$E_{ref,2}(x,y) = xyE_{f,MoS_2} + y(1-x)E_{f,MoTe_2} + x(1-y)E_{f,WS_2} + (1-x)(1-y)E_{f,WTe_2} \quad (5)$$

and

$$E_{ref,3}(x,y) = xyE_{f,MoSe_2} + y(1-x)E_{f,MoTe_2} + x(1-y)E_{f,WSe_2} + (1-x)(1-y)E_{f,WTe_2}. \quad (6)$$

We define the enthalpy of mixing ($E_{mix}$) by subtracting the reference energy from the enthalpy based on the DFT calculation of the quaternary system as,

$$E_{mix} = E_f(x,y) - E_{ref}(x,y) \quad (7)$$

In the calculation of energy of mixing for ternary 2D CCTMCAs such as MoSSe or MoWS$_2$, we use two binary alloys of MoS$_2$/MoSe$_2$, or MoS$_2$/WS$_2$ as the reference.

The configurational entropy of mixing of the system can be written as [150]

$$S_{mix}(x, y) = -k_B N[x \ln x + (1-x) \ln(1-x) + y \ln y + (1-y) \ln(1-y)] \qquad (8)$$

Based on Eq.8, we know that the nine ternary and three quaternary 2D CCTMCAs with $x = y = 0.5$ have the highest configurational entropy among all the 2D CCTMCAs. The Gibbs free energy of mixing ($G_{mix}$) for the 2D CCTMCAs system equals the energy of mixing ($E_{mix}$) subtracted by the multiplication of temperature ($T$) and configurational entropy ($S_{mix}$), i.e.,

$$G_{mix}^T(x, y) = E_{mix}(x, y) - TS_{mix}(x, y) \qquad (9)$$

In order to demonstrate the stability of the ternary 2D CCTMCAs, Table 6 lists the energies of mixing and Gibbs free energies of nine 2D CCTMCAs. Based on the relative contents of each element, the configuration entropy for those selected ternary and Janus ternary 2D CCTMCAs is $S = k_B \ln 2$ [150]. For the nine ternary 2D CCTMCAs, the negative energy of mixing $E_{mix}$ and the Gibbs free energy of MoWS$_2$, MoWSe$_2$, and MoWTe$_2$ indicates these three ternary alloys are stable at 0 K, 300 K and 600 K. For the six Janus structures, on the other hand, the energy of mixing at 0 K and 300 K are all positive, implying the unstable structures at 0 K and 300 K. MoSSe and WSSe become stable and have negative Gibbs free energies at 600 K. The Gibbs free energies of the Mo/W-S-Te and Mo/W-Se-Te based 2D CCTMCAs at 600 K remain positive, suggesting that even at high temperature, these four Janus structures are still unstable and likely to suffer from decomposition.

**Table 6**. The formation energies, Gibbs free energies of mixing of nine ternary monolayer 2D CCTMCAs with the 2$H$ structure.

|  | MoWS$_2$ | MoWSe$_2$ | MoWTe$_2$ |
| --- | --- | --- | --- |

| | | | | | | |
|---|---|---|---|---|---|---|
| $E_{\text{mix}}$ (eV) | -0.005 | | -0.004 | | -0.002 | |
| $G_{\text{mix}}^{300K}$ (eV) | -0.023 | | -0.022 | | -0.020 | |
| $G_{\text{mix}}^{600K}$ (eV) | -0.041 | | -0.040 | | -0.038 | |
| | MoSSe | WSSe | MoSTe | WSTe | MoSeTe | WSeTe |
| $E_{\text{mix}}$ (eV) | 0.027 | 0.030 | 0.199 | 0.224 | 0.078 | 0.088 |
| $G_{\text{mix}}^{300K}$ (eV) | 0.009 | 0.012 | 0.181 | 0.206 | 0.060 | 0.070 |
| $G_{\text{mix}}^{600K}$ (eV) | -0.009 | -0.006 | 0.163 | 0.188 | 0.042 | 0.052 |

**Table 7**. The formation energies, Gibbs free energies of mixing (at the temperatures of 300 K and 600 K), and nonideality of the four quaternary 2D CCTMCA monolayers with the 2*H* structure.

| | Mo$_{0.5}$W$_{0.5}$SSe | Mo$_{0.5}$W$_{0.5}$STe | Mo$_{0.5}$W$_{0.5}$SeTe |
|---|---|---|---|
| $E_{\text{mix}}$ (eV) | -0.003 | 0.052 | 0.028 |
| $G_{\text{mix}}^{300K}$ (eV) | -0.111 | -0.055 | -0.079 |
| $G_{\text{mix}}^{600K}$ (eV) | -0.218 | -0.163 | -0.187 |
| $\Delta\mu$ (eV) | 0.222 | 0.474 | 0.252 |

Table 7 summarizes the energy of mixing and Gibbs free energy of three quaternary 2D CCTMCAs at 0 K, 300 K, and 600 K. We can observe that the calculated formation energy of Mo$_{0.5}$W$_{0.5}$SSe is negative, indicating this quaternary 2D CCTMCA is stable at 0 K. Indeed, monolayer Mo$_{0.5}$W$_{0.5}$SSe has also been experimentally synthesized [151]. By contrast, for the other two 2D CCTMCAs, Mo$_{0.5}$W$_{0.5}$STe and Mo$_{0.5}$W$_{0.5}$SeTe, the formation energies are positive, which is related to the large lattice difference among the four corresponding binary TMCs of each quaternary 2D CCTMCA system [76,152]. The difference in the atomic radii of S and Te is larger than that between the atomic radii of Se and Te, which also explains the higher formation energy of $Mo_{0.5}W_{0.5}STe$ than $Mo_{0.5}W_{0.5}SeTe$.

However, it is possible to stabilize the quaternary systems by considering the temperature effect in the Gibbs free energy. For each of the three quaternary 2D CCTMCAs we choose, the configurational entropy is $3k_B\ln2$. Table 7 shows that the Gibbs free energies of mixing $G_{mix}$ for all the three 2D CCTMCAs are negative at room temperature, indicating the stable structures of three quaternary system at the temperature of 300 K and 600 K. We can see that the temperature-dependent entropy term contributes to the lowering of Gibbs free energy in a great deal, and both $Mo_{0.5}W_{0.5}STe$ and $Mo_{0.5}W_{0.5}SeTe$, which show the positive energy of mixing, have negative Gibbs free energies at 300 K. Compared to the Janus structure 2D CCTMCAs of Mo/W-S-Te and Mo/W-Se-Te, whose Gibbs free energies remain positive even at 600 K, we can conclude that the entropy effect is beneficial for stabilizing quaternary 2D CCTMCAs. Therefore, the method of designing multinary 2D CCTMCA not only result in the tunable properties, but also lead to stable alloying phase by taking the advantage of the entropy effect.

From Table 7 and the above discussion, we know that the positive formation energy of both $Mo_{0.5}W_{0.5}STe$ and $Mo_{0.5}W_{0.5}SeTe$ can be stabilized at high temperature. However, these two quaternary 2D CCTMCAs still show the unstable phase at low temperature, which could result in the phase separation. Even though the process of phase separation in quaternary 2D CCTCAs is complicated, we can predict the direction where the quaternary 2D CCTMCAs will undergo a phase separation into a group from two dissimilar binary TMCs. We assume that a quaternary 2D CCTMCA ($Mo_{0.5}W_{0.5}STe$, for example) can be made up from four different binary TMCs with different cation anion pairs ($MoS_2$, $WTe_2$, $MoTe_2$, and $WS_2$), which can further be categorized into two groups (($MoS_2$-$WTe_2$) and ($MoTe_2$-$WS_2$)). We then calculate the energy difference between these groups to predict the phase separation from quaternary 2D CCTMCA to binary TMCs group. As an example of the $Mo_{0.5}W_{0.5}STe$ phase transformation, by taking two groups of binary TMCs ($MoS_2$-$WTe_2$) and ($MoTe_2$-$WS_2$) as reference, we can predict the stable group when phase separation happens using the nonideality of the solution. The nonideality $\Delta\mu$ is calculated via the chemical potential difference between two groups of binary TMCs, such as $MoS_2$-$WTe_2$ and $MoTe_2$-$WS_2$ [16],

$$\Delta\mu_2 = \left(\mu_{MoS_2} + \mu_{WTe_2}\right) - \left(\mu_{MoTe_2} + \mu_{WS_2}\right) \tag{10}$$

Similarly, we write $\Delta\mu$ for the other two 2D CCTMCAs as

$$\Delta\mu_1 = \left(\mu_{MoS_2} + \mu_{WSe_2}\right) - \left(\mu_{MoSe_2} + \mu_{WS_2}\right) \tag{11}$$

and

$$\Delta\mu_3 = \left(\mu_{MoSe_2} + \mu_{WTe_2}\right) - \left(\mu_{MoTe_2} + \mu_{WSe_2}\right) \tag{12}$$

The calculated nonideality results for the three 2D CCTMCAs are shown in Table 7, where the $\Delta\mu_1$ is in agreement with the previous work [16]. The positive $\Delta\mu_2$ of $Mo_{0.5}W_{0.5}STe$ indicates that the latter binary group of ($MoTe_2$-$WS_2$) is more stable, into which the quaternary 2D CCTMCA would decompose. We can see from the results that for the three quaternary 2D CCTMCAs systems of Mo-W-S-Se, Mo-W-S-Te, and Mo-W-Se-Te, the most stable binary TMC groups are $MoSe_2$-$WS_2$, $MoTe_2$-$WS_2$, and $MoTe_2$-$WSe_2$, respectively. Previous experimental result has shown one example of the phase separation in Mo-W-S-Se quaternary 2D CCTMCAs in Ref. [16] as the spinodal decomposition of quaternary 2D CCTMCAs Mo-W-S-Se into two ternary 2D CCTMCAs within the miscibility gap.

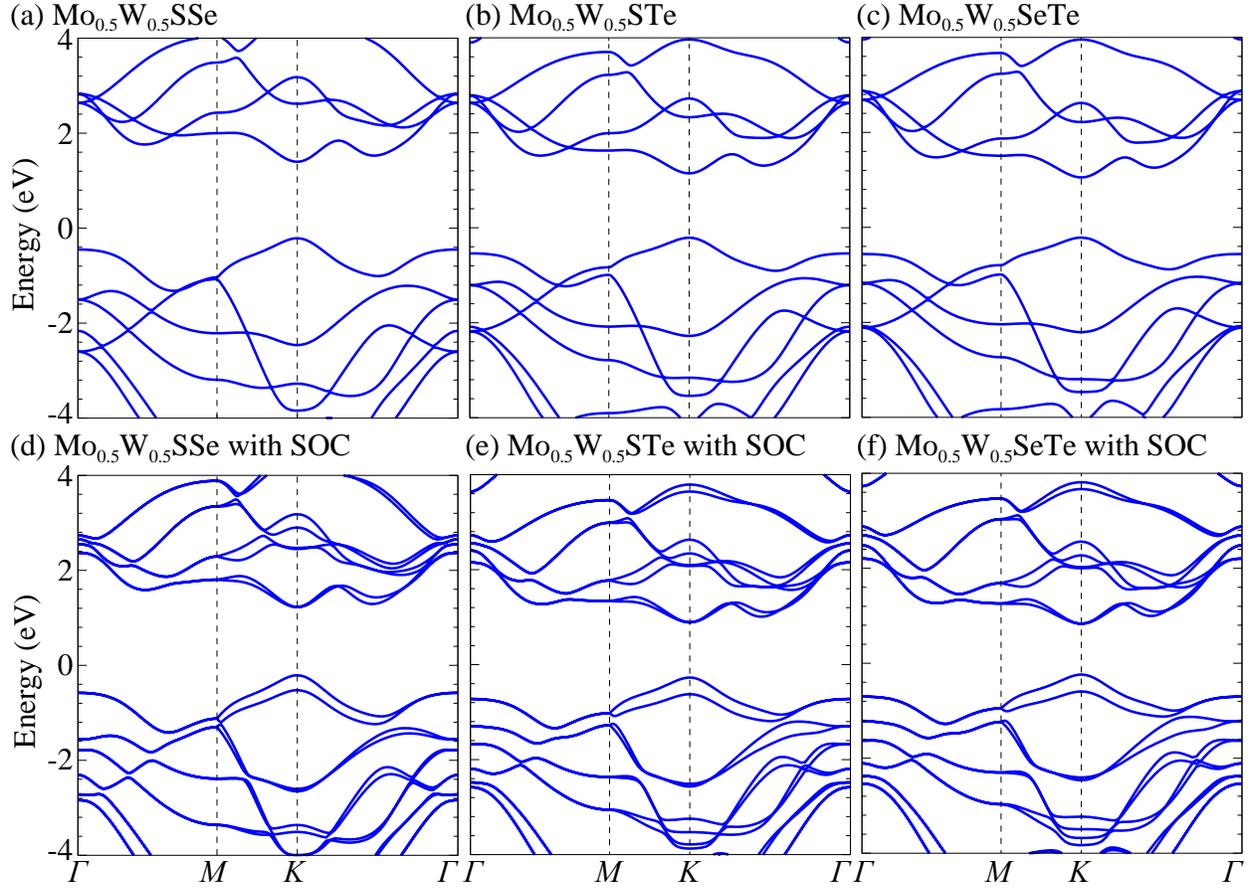

**Figure 10.** Band structures of three quaternary 2D CCTMCAs, Mo$_{0.5}$W$_{0.5}$SSe, Mo$_{0.5}$W$_{0.5}$STe, and Mo$_{0.5}$W$_{0.5}$SeTe. Spin-orbit coupling is not considered in the panels (a), (b), and (c), whereas it is accounted for in panels of (d), (e), and (f).

According to the calculated Gibbs free energies, we find that the three quaternary 2D CCTMCAs from SQS supercell are stable at room temperature. We henceforth focus on these three quaternary 2D CCTMCAs in discussing their electronic properties such as band structure and electrical conductivity, for the purpose of applying them as the potential materials in energy conversion applications. Figure 10 displays the band structures of three quaternary 2D CCTMCAs calculated with the PBE functional without and with considering the spin-orbit coupling (SOC). We can see that the calculations with or without considering SOC show the direct band gap of these three 2D CCTMCAs at $K$ point. The SO splitting energies at the $K$ point for $Mo_{0.5}W_{0.5}SSe$, $Mo_{0.5}W_{0.5}STe$, and $Mo_{0.5}W_{0.5}SeTe$ are 0.31, 0.35, and 0.36 eV, respectively.

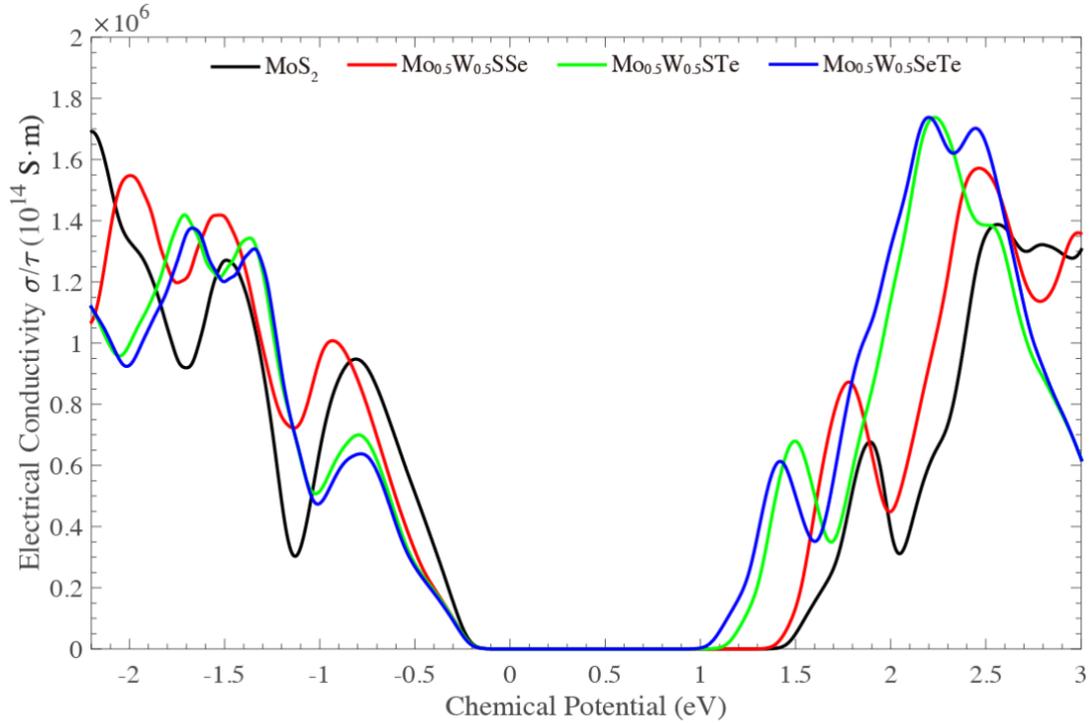

**Figure 11**. Electrical conductivity of $Mo_{0.5}W_{0.5}SSe$, $Mo_{0.5}W_{0.5}STe$, and $Mo_{0.5}W_{0.5}SeTe$ at 300 K as a function of the chemical potential. Electrical conductivity of $MoS_2$ is also plotted as benchmark.

Figure 11 displays the electrical conductivity of $Mo_{0.5}W_{0.5}SSe$, $Mo_{0.5}W_{0.5}STe$, and $Mo_{0.5}W_{0.5}SeTe$ at 300 K. We also compute the electrical conductivity of monolayer $MoS_2$ and benchmark the results with the literature [153]. The relaxation time used in these calculations is taken as an approximated constant value of 10.0 fs, which has been used in the calculations of electrical conductivity for other monolayer semiconductors such as SnSe, $Sc_2C$, and $TiSe_2$ [154-156]. We first observe that the band gaps of these three 2D CCTMCAs, corresponding to the regions where the conductivity equals to zero, are in agreement with the VCA-DFT results (see Figure 5). These band gaps lie from the visible light range to the near-infrared region, which enable the photovoltaic effect in a wider region to enhance the photovoltaic conversion efficiency [157].

In Figure 11, the negative (left) side of the chemical potential illustrates the holes conductivity (*p*-type), and the positive (right) side is for the conductivity from electrons (*n*-type). By changing the chemical potential that can be realized by different methods such as doping or applied gate voltage [158], 2D CCTMCAs can reach a high conductivity of $1.7 \times 10_6$ S·m, which guarantees the high carrier transport within the single layer. Moreover, the three quaternary 2D CCTMCAs exhibit higher electron conductivity comparing to pristine $MoS_2$ monolayer. The high conductivity of three examples shows their potential in the energy conversion applications such as photovoltaics.

**Conclusions**

In summary, we propose a workflow using the DFT calculation from VCA models in search of suitable multinary 2D CCTMCAs in different applications in energy conversion and spintronics. We have computationally characterized five critical structural and electrical properties of 2D CCTMCAs and also benchmarked DFT results using unit cell and SQS models to validate the accuracy of calculation from VCA method. We find that VCA-DFT calculations lead to comparable results of lattice constants, band gap, electron and hole effective masses, spin orbit splitting, and band alignment with the unit-cell-DFT and SQS-DFT results, with some exceptions in the CBM, effective hole masses, and band gaps of MoSTe and WSTe, which are caused by the inaccurate prediction of the location of CBM in VCA-DFT calculations. Our results show that the multinary 2D CCTMCAs exhibit tunable properties such as band gaps, lattice constants, effective masses, and band alignments. These tunable properties are helpful in designing the lattice matching type II heterostructures for the applications of light adsorption and conversion devices. The strong SOC of 2D CCTMCAs also suggest the possibility of utilizing the multinary 2D CCTMCAs in spintronics. In addition to the high-throughput computational characterization of 2D CCTMCAs workflow, we propose three quaternary 2D CCTMCAs, $Mo_{0.5}W_{0.5}SSe$, $Mo_{0.5}W_{0.5}STe$, and Mo$_{0.5}$W$_{0.5}$SeTe at room temperature, serving as excellent examples to illustrate the entropy-stabilized alloys from multiple component design of 2D CCTMCAs. In addition, they also show high electrical conductivity as promising materials for energy conversion applications. Although currently most research on quaternary TMC alloys centers on Mo/W-based alloys, 2D nanosheets of TaSe$_2$, NbSe$_2$, and NiTe$_2$ (Ta: Tantalum, Nb: Niobium) have been obtained in experiments, indicating that alloying of these TMCs may be used to develop 2D CCTMCAs [159]. The high-throughput workflow we proposed enables the extension of the research on the 2D CCTMCAs consisting of other transition metal, metal, and chalcogen elements.

## Acknowledgements

We thank the start-up funds from Arizona State University (ASU). N. B. thanks the Science and Engineering Experience (SCENE) program as ASU. This research used computational resources of the Texas Advanced Computing Center under Contracts No.TG-DMR170070 and the Agave cluster at ASU.

## References

[1]     S. Manzeli, D. Ovchinnikov, D. Pasquier, O. V. Yazyev, and A. Kis, 2D transition metal dichalcogenides, Nature Reviews Materials **2**, 17033 (2017).
[2]     Z. Lei, X. Liu, H. Wang, Y. Wu, S. Jiang, and Z. Lu, Development of advanced materials via entropy engineering, Scripta Materialia **165**, 164 (2019).
[3]     Y. Ding, Y. Wang, J. Ni, L. Shi, S. Shi, and W. Tang, First principles study of structural, vibrational and electronic properties of graphene-like *MX*$_2$ (*M*= Mo, Nb, W, Ta; *X*= S, Se, Te) monolayers, Physica B: Condensed Matter **406**, 2254 (2011).
[4]     J. Chang, L. F. Register, and S. K. Banerjee, Ballistic performance comparison of monolayer transition metal dichalcogenide *MX*$_2$ (*M*= Mo, W; *X*= S, Se, Te) MOSFETs, Journal of Applied Physics  (2014).
[5]     M. M. Ugeda *et al.*, Giant bandgap renormalization and excitonic effects in a monolayer transition metal dichalcogenide semiconductor, Nature Materials **13**, 1091 (2014).
[6]     B. Radisavljevic, A. Radenovic, J. Brivio, V. Giacometti, and A. Kis, Single-layer MoS$_2$ transistors, Nature Nanotechnology **6**, 147 (2011).
[7]     B. Ozdemir and V. Barone, Thickness dependence of solar cell efficiency in transition metal dichalcogenides *MX*$_2$ (*M*: Mo, W; *X*: S, Se, Te), Solar Energy Materials and Solar Cells **212**, 110557 (2020).

[8] A. Kumar and P. Ahluwalia, Electronic structure of transition metal dichalcogenides monolayers 1*H-MX*$_2$ (*M*= Mo, W; *X*= S, Se, Te) from ab-initio theory: new direct band gap semiconductors, The European Physical Journal B **85**, 1 (2012).
[9] S. Susarla *et al.*, Quaternary 2D transition metal dichalcogenides (TMDs) with tunable bandgap, Advanced Materials **29**, 1702457 (2017).
[10] V. Kochat *et al.*, Re doping in 2D transition metal dichalcogenides as a new route to tailor structural phases and induced magnetism, Advanced Materials **29**, 1703754 (2017).
[11] Z. Guan, S. Ni, and S. Hu, Tunable electronic and optical properties of monolayer and multilayer Janus MoSSe as a photocatalyst for solar water splitting: a first-principles study, The Journal of Physical Chemistry C **122**, 6209 (2018).
[12] C. Gong *et al.*, 2D nanomaterial arrays for electronics and optoelectronics, Advanced Functional Materials **28**, 1706559 (2018).
[13] Y. Liu, X. Duan, Y. Huang, and X. Duan, Two-dimensional transistors beyond graphene and TMDCs, Chemical Society Reviews **47**, 6388 (2018).
[14] W. S. Yun, S. Han, S. C. Hong, I. G. Kim, and J. Lee, Thickness and strain effects on electronic structures of transition metal dichalcogenides: 2H-*MX*$_2$ semiconductors (*M*= Mo, W; *X*= S, Se, Te), Physical Review B **85**, 033305 (2012).
[15] P. Johari and V. B. Shenoy, Tuning the electronic properties of semiconducting transition metal dichalcogenides by applying mechanical strains, ACS Nano **6**, 5449 (2012).
[16] S. Susarla *et al.*, Quaternary alloys: Thermally induced 2D alloy-heterostructure transformation in quaternary alloys, Advanced Materials **30**, 1870344 (2018).
[17] T. Jakubczyk, V. Delmonte, M. Koperski, K. Nogajewski, C. Faugeras, W. Langbein, M. Potemski, and J. Kasprzak, Radiatively limited dephasing and exciton dynamics in MoSe$_2$ monolayers revealed with four-wave mixing microscopy, Nano Letters **16**, 5333 (2016).
[18] L. Britnell *et al.*, Field-effect tunneling transistor based on vertical graphene heterostructures, Science **335**, 947 (2012).
[19] X. Li *et al.*, 18.5% efficient graphene/GaAs van der Waals heterostructure solar cell, Nano Energy **16**, 310 (2015).
[20] Y. Tan, X. Liu, Z. He, Y. Liu, M. Zhao, H. Zhang, and F. Chen, Tuning of interlayer coupling in large-area graphene/WSe$_2$ van der Waals heterostructure via ion irradiation: optical evidences and photonic applications, ACS Photonics **4**, 1531 (2017).
[21] K. Wu, H. Ma, Y. Gao, W. Hu, and J. Yang, Highly-efficient heterojunction solar cells based on two-dimensional tellurene and transition metal dichalcogenides, Journal of Materials Chemistry A **7**, 7430 (2019).
[22] X. Duan *et al.*, Lateral epitaxial growth of two-dimensional layered semiconductor heterojunctions, Nature Nanotechnology **9**, 1024 (2014).
[23] W. Wei, Y. Dai, and B. Huang, In-plane interfacing effects of two-dimensional transition-metal dichalcogenide heterostructures, Physical Chemistry Chemical Physics **18**, 15632 (2016).
[24] B. Amin, N. Singh, and U. Schwingenschlögl, Heterostructures of transition metal dichalcogenides, Physical Review B **92**, 075439 (2015).
[25] C. Mu, W. Wei, J. Li, B. Huang, and Y. Dai, Electronic properties of two-dimensional in-plane heterostructures of WS$_2$/WSe$_2$/MoS$_2$, Materials Research Express **5**, 046307 (2018).
[26] Y. Gong *et al.*, Vertical and in-plane heterostructures from WS$_2$/MoS$_2$ monolayers, Nature Materials **13**, 1135 (2014).

[27] E. Abbasi and K. Dehghani, Hot tensile properties of CoCrFeMnNi (NbC) compositionally complex alloys, Materials Science and Engineering: A **772**, 138771 (2020).

[28] E. Abbasi and K. Dehghani, Microstructure and mechanical properties of $Co_{19}Cr_{20}Fe_{20}Mn_{21}Ni_{19}$ and $Co_{19}Cr_{20}Fe_{20}Mn_{21}Ni_{119}Nb_{0.06}C_{0.8}$ high-entropy/compositionally-complex alloys after annealing, Materials Science and Engineering: A **772**, 138812 (2020).

[29] A. M. Manzoni and U. Glatzel, New multiphase compositionally complex alloys driven by the high entropy alloy approach, Materials Characterization **147**, 512 (2019).

[30] T. M. Butler and M. L. Weaver, Oxidation behavior of arc melted AlCoCrFeNi multi-component high-entropy alloys, Journal of Alloys and Compounds **674**, 229 (2016).

[31] K. Jin and H. Bei, Single-phase concentrated solid-solution alloys: Bridging intrinsic transport properties and irradiation resistance, Frontiers in Materials **5**, 26 (2018).

[32] O. N. Senkov, J. D. Miller, D. B. Miracle, and C. Woodward, Accelerated exploration of multi-principal element alloys with solid solution phases, Nature Communications **6**, 6529 (2015).

[33] Y. Zhang, T. Zuo, Y. Cheng, and P. K. Liaw, High-entropy alloys with high saturation magnetization, electrical resistivity, and malleability, Scientific Reports **3**, 1455 (2013).

[34] O. N. Senkov, G. Wilks, J. Scott, and D. B. Miracle, Mechanical properties of $Nb_{25}Mo_{25}Ta_{25}W_{25}$ and $V_{20}Nb_{20}Mo_{20}Ta_{20}W_{20}$ refractory high entropy alloys, Intermetallics **19**, 698 (2011).

[35] W. Guo, W. Dmowski, J.-Y. Noh, P. Rack, P. K. Liaw, and T. Egami, Local atomic structure of a high-entropy alloy: An X-ray and neutron scattering study, Metallurgical Materials Transactions A **44**, 1994 (2013).

[36] C.-C. Tung, J.-W. Yeh, T.-t. Shun, S.-K. Chen, Y.-S. Huang, and H.-C. Chen, On the elemental effect of AlCoCrCuFeNi high-entropy alloy system, Materials Letters **61**, 1 (2007).

[37] Y. Zhang *et al.*, Influence of chemical disorder on energy dissipation and defect evolution in concentrated solid solution alloys, Nature Communications **6**, 8736 (2015).

[38] H. S. Oh *et al.*, Engineering atomic-level complexity in high-entropy and complex concentrated alloys, Nature Communications **10**, 1 (2019).

[39] T. Zuo *et al.*, Tailoring magnetic behavior of CoFeMnNi*X* (*X*= Al, Cr, Ga, and Sn) high entropy alloys by metal doping, Acta Materialia **130**, 10 (2017).

[40] S. Shafeie, S. Guo, Q. Hu, H. Fahlquist, P. Erhart, and A. Palmqvist, High-entropy alloys as high-temperature thermoelectric materials, Journal of Applied Physics **118**, 184905 (2015).

[41] Z. Fan, H. Wang, Y. Wu, X. Liu, and Z. Lu, Thermoelectric performance of PbSnTeSe high-entropy alloys, Materials Research Letters **5**, 1 (2016).

[42] Z. Fan, H. Wang, Y. Wu, X. Liu, and Z. Lu, Thermoelectric high-entropy alloys with low lattice thermal conductivity, RSC Advances **6**, 52164 (2016).

[43] K. Jin, B. C. Sales, G. M. Stocks, G. D. Samolyuk, M. Daene, W. J. Weber, Y. Zhang, and H. Bei, Tailoring the physical properties of Ni-based single-phase equiatomic alloys by modifying the chemical complexity, Scientific Reports **6**, 20159 (2016).

[44] H. Watzinger, J. Kukučka, L. Vukušić, F. Gao, T. Wang, F. Schäffler, J.-J. Zhang, and G. Katsaros, A germanium hole spin qubit, Nature Communications **9**, 3902 (2018).

[45] A. Perrin, M. Sorescu, M.-T. Burton, D. E. Laughlin, and M. McHenry, The role of compositional tuning of the distributed exchange on magnetocaloric properties of high-entropy alloys, JOM **69**, 2125 (2017).

[46] H. Kato, S. Adachi, H. Nakanishi, and K. Ohtsuka, Optical properties of $(Al_xGa_{1-x})_{0.5}In_{0.5}P$ quaternary alloys, Japanese journal of applied physics **33**, 186 (1994).

[47]　G. M. Ford, Q. Guo, R. Agrawal, and H. W. Hillhouse, Earth abundant element Cu$_2$Zn(Sn$_{1-x}$Ge$_x$)S$_4$ nanocrystals for tunable band gap solar cells: 6.8% efficient device fabrication, Chemistry of Materials **23**, 2626 (2011).

[48]　J. He, L. Sun, S. Chen, Y. Chen, P. Yang, and J. Chu, Composition dependence of structure and optical properties of Cu$_2$ZnSn(S, Se)$_4$ solid solutions: an experimental study, Journal of Alloys and Compounds **511**, 129 (2012).

[49]　C. Tan, Z. Lai, and H. Zhang, Ultrathin two-dimensional multinary layered metal chalcogenide nanomaterials, Advanced Materials **29**, 1701392 (2017).

[50]　H. Cotal, C. Fetzer, J. Boisvert, G. Kinsey, R. King, P. Hebert, H. Yoon, and N. Karam, III–V multijunction solar cells for concentrating photovoltaics, Energy Environmental Science **2**, 174 (2009).

[51]　S. H. Su, Y. T. Hsu, Y. H. Chang, M. H. Chiu, C. L. Hsu, W. T. Hsu, W. H. Chang, J. H. He, and L. J. Li, Band gap-tunable molybdenum sulfide selenide monolayer alloy, Small **10**, 2589 (2014).

[52]　P. Zawadzki, A. Zakutayev, and S. Lany, Entropy-driven clustering in tetrahedrally bonded multinary materials, Physical Review Applied **3**, 034007 (2015).

[53]　E. Torun, H. Sahin, S. Cahangirov, A. Rubio, and F. Peeters, Anisotropic electronic, mechanical, and optical properties of monolayer WTe$_2$, Journal of Applied Physics **119**, 074307 (2016).

[54]　X. Zhang, X. Shi, W. Ye, C. Ma, and C. Wang, Electrochemical deposition of quaternary Cu$_2$ZnSnS$_4$ thin films as potential solar cell material, Applied Physics A **94**, 381 (2009).

[55]　R. J. Walters *et al.*, in *2011 37th IEEE Photovoltaic Specialists Conference* (IEEE, 2011), pp. 000122.

[56]　F. Özel, A. Sarılmaz, B. İstanbullu, A. Aljabour, M. Kuş, and S. Sönmezoğlu, Penternary chalcogenides nanocrystals as catalytic materials for efficient counter electrodes in dye-synthesized solar cells, Scientific Reports **6**, 1 (2016).

[57]　F. Ceballos, M. Z. Bellus, H. Y. Chiu, and H. Zhao, Ultrafast charge separation and indirect exciton formation in a MoS$_2$-MoSe$_2$ van der Waals heterostructure, ACS Nano **8**, 12717 (2014).

[58]　L. Li, Growth mode and characterization of Si/SiC heterostructure of large lattice-mismatch, Heterojunctions and Nanostructures, 67 (2018).

[59]　A. Bhattacharyya and D. Maurice, On the evolution of stresses due to lattice misfit at a Ni-superalloy and YSZ interface, Surfaces and Interfaces **12**, 86 (2018).

[60]　L. Ju, M. Bie, X. Tang, J. Shang, and L. Kou, Janus WSSe Monolayer: Excellent Photocatalyst for Overall Water-splitting, ACS Applied Materials & Interfaces (2020).

[61]　Q.-F. Yao, J. Cai, W.-Y. Tong, S.-J. Gong, J.-Q. Wang, X. Wan, C.-G. Duan, and J. Chu, Manipulation of the large Rashba spin splitting in polar two-dimensional transition-metal dichalcogenides, Physical Review B **95**, 165401 (2017).

[62]　A. P. Tiwari, T. G. Novak, X. Bu, J. C. Ho, and S. Jeon, Layered ternary and quaternary transition metal chalcogenide based catalysts for water splitting, Catalysts **8**, 551 (2018).

[63]　Y. Cheng, Z. Zhu, M. Tahir, and U. Schwingenschlögl, Spin-orbit–induced spin splittings in polar transition metal dichalcogenide monolayers, EPL (Europhysics Letters) **102**, 57001 (2013).

[64]　Y. Zhang, H. Ye, Z. Yu, Y. Liu, and Y. Li, First-principles study of square phase M$X_2$ and Janus M$XY$ (M= Mo, W; $X$, $Y$= S, Se, Te) transition metal dichalcogenide monolayers under biaxial strain, Physica E: Low-dimensional Systems and Nanostructures **110**, 134 (2019).

[65] X. J. Wu, X. Huang, X. Qi, H. Li, B. Li, and H. Zhang, Copper-based ternary and quaternary semiconductor nanoplates: Templated synthesis, characterization, and photoelectrochemical properties, Angewandte Chemie International Edition **53**, 8929 (2014).

[66] M. C. Johnson, C. Wrasman, X. Zhang, M. Manno, C. Leighton, and E. S. Aydil, Self-regulation of Cu/Sn ratio in the synthesis of $Cu_2ZnSnS_4$ films, Chemistry of Materials **27**, 2507 (2015).

[67] J. B. Varley, X. He, A. Rockett, and V. Lordi, Stability of $Cd_{1-x}Zn_xO_yS_{1-y}$ Quaternary Alloys Assessed with First-Principles Calculations, ACS Applied Materials & Interfaces **9**, 5673 (2017).

[68] Y. Liang, J. Li, H. Jin, B. Huang, and Y. Dai, Photoexcitation dynamics in Janus-$MoSSe/WSe_2$ heterobilayers: ab initio time-domain study, The Journal of Physical Chemistry Letters **9**, 2797 (2018).

[69] M. Idrees, H. Din, R. Ali, G. Rehman, T. Hussain, C. Nguyen, I. Ahmad, and B. Amin, Optoelectronic and solar cell applications of Janus monolayers and their van der Waals heterostructures, Physical Chemistry Chemical Physics **21**, 18612 (2019).

[70] B. K. Ridley, *Quantum processes in semiconductors* (Oxford University Press, B. K. Ridley, 2013).

[71] T. Kirchartz and U. Rau, Linking structural properties with functionality in solar cell materials–the effective mass and effective density of states, Sustainable Energy & Fuels **2**, 1550 (2018).

[72] A. P. Tiwari, D. Kim, Y. Kim, O. Prakash, and H. Lee, Highly active and stable layered ternary transition metal chalcogenide for hydrogen evolution reaction, Nano Energy **28**, 366 (2016).

[73] H. L. Zhuang and R. G. Hennig, Computational search for single-layer transition-metal dichalcogenide photocatalysts, The Journal of Physical Chemistry C **117**, 20440 (2013).

[74] B. Schuler *et al.*, Large spin-orbit splitting of deep in-gap defect states of engineered sulfur vacancies in monolayer $WS_2$, Physical Review Letters **123**, 076801 (2019).

[75] D. S. Abergel, J. M. Edge, and A. V. Balatsky, The role of spin–orbit coupling in topologically protected interface states in Dirac materials, New Journal of Physics **16**, 065012 (2014).

[76] J. Kang, S. Tongay, J. Li, and J. Wu, Monolayer semiconducting transition metal dichalcogenide alloys: Stability and band bowing, Journal of Applied Physics **113**, 143703 (2013).

[77] S. Mankovsky, K. Chadova, D. Koedderitzsch, J. Minár, H. Ebert, and W. Bensch, Electronic, magnetic, and transport properties of Fe-intercalated $2H$-$TaS_2$ studied by means of the KKR-CPA method, Physical Review B **92**, 144413 (2015).

[78] D. D. Johnson, D. Nicholson, F. Pinski, B. Gyorffy, and G. Stocks, Density-functional theory for random alloys: total energy within the coherent-potential approximation, Physical Review Letters **56**, 2088 (1986).

[79] J. Faulkner, N. Moghadam, Y. Wang, and G. M. Stocks, Comparison of the electronic states of alloys from the coherent potential approximation and an order-N calculation, Journal of phase equilibria **19**, 538 (1998).

[80] S. Mu, S. Wimmer, S. Mankovsky, H. Ebert, and G. Stocks, Influence of local lattice distortions on electrical transport of refractory high entropy alloys, Scripta Materialia **170**, 189 (2019).

[81] K. Wood and J. B. Pendry, Layer Method for Band Structure of Layer Compounds, Physical Review Letters **31**, 1400 (1973).

[82] F. Tian, A review of solid-solution models of high-entropy alloys based on ab initio calculations, Frontiers in Materials **4**, 36 (2017).
[83] X. Qian, P. Jiang, P. Yu, X. Gu, Z. Liu, and R. Yang, Anisotropic thermal transport in van der Waals layered alloys WSe$_{2(1-x)}$Te$_{2x}$, Applied Physics Letters **112**, 241901 (2018).
[84] U.-G. Jong, C.-J. Yu, J.-S. Ri, N.-H. Kim, and G.-C. Ri, Influence of halide composition on the structural, electronic, and optical properties of mixed CH$_3$NH$_3$Pb(I$_{1-x}$Br$_x$)$_3$ perovskites calculated using the virtual crystal approximation method, Physical Review B **94**, 125139 (2016).
[85] L. Bellaiche and D. Vanderbilt, Virtual crystal approximation revisited: Application to dielectric and piezoelectric properties of perovskites, Physical Review B **61**, 7877 (2000).
[86] N. J. Ramer and A. M. Rappe, Virtual-crystal approximation that works: locating a compositional phase boundary in Pb(Zr$_{1-x}$Ti$_x$)O$_3$, Physical Review B **62**, R743 (2000).
[87] S. A.-B. Nasrallah, S. B. Afia, H. Belmabrouk, and M. Said, Optoelectronic properties of zinc blende ZnSSe and ZnBeTe alloys, The European Physical Journal B **43**, 3 (2005).
[88] C. Tablero, Electronic and optical property analysis of the Cu-Sb-S tetrahedrites for high-efficiency absorption devices, The Journal of Physical Chemistry C **118**, 15122 (2014).
[89] D. J. Wilson, B. Winkler, E. A. Juarez-Arellano, A. Friedrich, K. Knorr, C. J. Pickard, and V. Milman, Virtual crystal approximation study of nitridosilicates and oxonitridoaluminosilicates, Journal of Physics and Chemistry of Solids **69**, 1861 (2008).
[90] G. Hua and D. Li, A first-principles study on the mechanical and thermodynamic properties of (Nb$_{1-x}$Ti$_x$)C complex carbides based on virtual crystal approximation, RSC Advances **5**, 103686 (2015).
[91] G. Kresse and J. Furthmüller, Efficient iterative schemes for ab initio total-energy calculations using a plane-wave basis set, Physical Review B **54**, 11169 (1996).
[92] A. van de Walle, M. Asta, and G. Ceder, The alloy theoretic automated toolkit: A user guide, Calphad (2002).
[93] P. E. Blöchl, Projector augmented-wave method, Physical Review B **50**, 17953 (1994).
[94] G. Kresse and D. Joubert, From ultrasoft pseudopotentials to the projector augmented-wave method, Physical Review B **59**, 1758 (1999).
[95] J. P. Perdew, K. Burke, and M. Ernzerhof, Generalized gradient approximation made simple, Physical Review Letters **77**, 3865 (1996).
[96] D. O. Scanlon and G. W. Watson, Band gap anomalies of the ZnM$_2^{III}$O$_4$ (M$_{III}$= Co, Rh, Ir) spinels, Physical Chemistry Chemical Physics **13**, 9667 (2011).
[97] F. Tran, P. Blaha, and K. Schwarz, Band gap calculations with Becke-Johnson exchange potential, Journal of Physics: Condensed Matter **19**, 196208 (2007).
[98] H. Li *et al.*, Anomalous behavior of 2d janus excitonic layers under extreme pressures, Advanced Materials, 2002401 (2020).
[99] H. J. Monkhorst and J. D. Pack, Special points for Brillouin-zone integrations, Physical Review B **13**, 5188 (1976).
[100] A. Ganose, A. Jackson, and D. Scanlon, sumo: Command-line tools for plotting and analysis of periodic *ab initio* calculations, Journal of Open Source Software **3**, 717 (2018).
[101] G. K. Madsen and D. J. Singh, BoltzTraP. A code for calculating band-structure dependent quantities, Computer Physics Communications **175**, 67 (2006).
[102] S. P. Ong *et al.*, Python Materials Genomics (pymatgen): A robust, open-source python library for materials analysis, Computational Materials Science **68**, 314 (2013).

[103] L. D. Whalley, J. M. Frost, B. J. Morgan, and A. Walsh, Impact of nonparabolic electronic band structure on the optical and transport properties of photovoltaic materials, Physical Review B **99**, 085207 (2019).

[104] D. M. Riffe, Temperature dependence of silicon carrier effective masses with application to femtosecond reflectivity measurements, JOSA B **19**, 1092 (2002).

[105] J. R. Schaibley, H. Yu, G. Clark, P. Rivera, J. S. Ross, K. L. Seyler, W. Yao, and X. Xu, Valleytronics in 2D materials, Nature Reviews Materials **1**, 1 (2016).

[106] K. F. Mak, K. He, J. Shan, and T. F. Heinz, Control of valley polarization in monolayer $MoS_2$ by optical helicity, Nature Nanotechnology **7**, 494 (2012).

[107] J. Yan *et al.*, Stacking-dependent interlayer coupling in trilayer $MoS_2$ with broken inversion symmetry, Nano Letters **15**, 8155 (2015).

[108] Y. Liu, Y. Gao, S. Zhang, J. He, J. Yu, and Z. Liu, Valleytronics in transition metal dichalcogenides materials, Nano Research **12**, 2695 (2019).

[109] R. Herberholz, V. Nadenau, U. Rühle, C. Köble, H. Schock, and B. Dimmler, Prospects of wide-gap chalcopyrites for thin film photovoltaic modules, Solar Energy Materials and Solar Cells **49**, 227 (1997).

[110] T. Minemoto, Y. Hashimoto, T. Satoh, T. Negami, H. Takakura, and Y. Hamakawa, Cu(In, Ga)$Se_2$ solar cells with controlled conduction band offset of window/Cu(In, Ga)$Se_2$ layers, Journal of Applied Physics **89**, 8327 (2001).

[111] C. Li, Q. Cao, F. Wang, Y. Xiao, Y. Li, J. J. Delaunay, and H. Zhu, Engineering graphene and TMDs based van der Waals heterostructures for photovoltaic and photoelectrochemical solar energy conversion, Chemical Society Reviews **47**, 4981 (2018).

[112] U. Dasgupta, A. Bera, and A. J. Pal, Band diagram of heterojunction solar cells through scanning tunneling spectroscopy, ACS Energy Letters **2**, 582 (2017).

[113] M. Xie, B. Cai, Z. Meng, Y. Gu, S. Zhang, X. Liu, L. Gong, X. Li, and H. Zeng, Two-dimensional BaS/InTe: A promising tandem solar cell with high power conversion efficiency, ACS Appl Mater Interfaces **12**, 6074 (2020).

[114] N. Wang, D. Cao, J. Wang, P. Liang, X. Chen, and H. Shu, Interface effect on electronic and optical properties of antimonene/GaAs van der Waals heterostructures, Journal of Materials Chemistry C **5**, 9687 (2017).

[115] C. Espejo, T. Rangel, A. Romero, X. Gonze, and G.-M. Rignanese, Band structure tunability in $MoS_2$ under interlayer compression: A DFT and GW study, Physical Review B **87**, 245114 (2013).

[116] S. Bhattacharyya and A. K. Singh, Semiconductor-metal transition in semiconducting bilayer sheets of transition-metal dichalcogenides, Physical Review B **86**, 075454 (2012).

[117] R. D. Shannon, Revised effective ionic radii and systematic studies of interatomic distances in halides and chalcogenides, Acta Crystallographica Section A **32**, 751 (1976).

[118] B. Cordero, V. Gómez, A. E. Platero-Prats, M. Revés, J. Echeverría, E. Cremades, F. Barragán, and S. Alvarez, Covalent radii revisited, Dalton Transactions, 2832 (2008).

[119] C. Gong, H. Zhang, W. Wang, L. Colombo, R. M. Wallace, and K. Cho, Band alignment of two-dimensional transition metal dichalcogenides: Application in tunnel field effect transistors, Applied Physics Letters **103**, 053513 (2013).

[120] M. Kan, H. G. Nam, Y. H. Lee, and Q. Sun, Phase stability and Raman vibration of the molybdenum ditelluride ($MoTe_2$) monolayer, Physical Chemistry Chemical Physics **17**, 14866 (2015).

[121] H. Huang, X. Fan, D. J. Singh, H. Chen, Q. Jiang, and W. Zheng, Controlling phase transition for single-layer MTe$_2$ (M= Mo and W): modulation of the potential barrier under strain, Physical Chemistry Chemical Physics **18**, 4086 (2016).

[122] C. Zhang *et al.*, Systematic study of electronic structure and band alignment of monolayer transition metal dichalcogenides in Van der Waals heterostructures, 2D Materials **4**, 015026 (2016).

[123] Z. Jin, X. Li, J. T. Mullen, and K. W. Kim, Intrinsic transport properties of electrons and holes in monolayer transition-metal dichalcogenides, Physical Review B **90**, 045422 (2014).

[124] D. Rhodes, S. Das, Q. R. Zhang, B. Zeng, N. Pradhan, N. Kikugawa, E. Manousakis, and L. Balicas, Role of spin-orbit coupling and evolution of the electronic structure of WTe$_2$ under an external magnetic field, Physical Review B **92**, 125152 (2015).

[125] J. Kang, S. Tongay, J. Zhou, J. Li, and J. Wu, Band offsets and heterostructures of two-dimensional semiconductors, Applied Physics Letters **102**, 012111 (2013).

[126] Y. Liang, S. Huang, R. Soklaski, and L. Yang, Quasiparticle band-edge energy and band offsets of monolayer of molybdenum and tungsten chalcogenides, Applied Physics Letters **103**, 042106 (2013).

[127] P. M. Gehring, Neutron diffuse scattering in lead-based relaxor ferroelectrics and its relationship to the ultra-high piezoelectricity, Journal of Advanced Dielectrics **2**, 1241005 (2012).

[128] K. Leung, E. Cockayne, and A. Wright, Effective Hamiltonian study of PbZr$_{0.95}$Ti$_{0.05}$O$_3$, Physical Review B **65**, 214111 (2002).

[129] A. Poursaleh and A. Andalib, An all optical majority gate using nonlinear photonic crystal based ring resonators, Optica Applicata **49** (2019).

[130] T. Dargam, R. Capaz, and B. Koiller, Critical analysis of the virtual crystal approximation, Brazilian Journal of Physics **27**, 299 (1997).

[131] C. Chen, E. Wang, Y. Gu, D. Bylander, and L. Kleinman, Unexpected band-gap collapse in quaternary alloys at the group-III-nitride/GaAs interface: GaAlAsN, Physical Review B **57**, 3753 (1998).

[132] L. Vegard, Die konstitution der mischkristalle und die raumfüllung der atome, Zeitschrift für Physik **5**, 17 (1921).

[133] Y. Luo, K. Ren, S. Wang, J.-P. Chou, J. Yu, Z. Sun, and M. Sun, First-principles study on transition-metal dichalcogenide/BSe van der Waals heterostructures: A promising water-splitting photocatalyst, The Journal of Physical Chemistry C **123**, 22742 (2019).

[134] Q. H. Wang, K. Kalantar-Zadeh, A. Kis, J. N. Coleman, and M. S. Strano, Electronics and optoelectronics of two-dimensional transition metal dichalcogenides, Nature Nanotechnology **7**, 699 (2012).

[135] M. G. Ju, J. Dai, L. Ma, and X. C. Zeng, Perovskite chalcogenides with optimal bandgap and desired optical absorption for photovoltaic devices, Advanced Energy Materials **7**, 1700216 (2017).

[136] H. I. Eya, E. Ntsoenzok, and N. Y. Dzade, First–Principles Investigation of the Structural, Elastic, Electronic, and Optical Properties of α–and β–SrZrS$_3$: Implications for Photovoltaic Applications, Materials **13**, 978 (2020).

[137] T. Mueller, TMD-based photodetectors, light emitters and photovoltaics, 2D Materials for Nanoelectronics **17**, 13 (2016).

[138] D. Huertas-Hernando, F. Guinea, and A. Brataas, Spin-orbit coupling in curved graphene, fullerenes, nanotubes, and nanotube caps, Physical Review B **74**, 155426 (2006).

[139] S. Bhatti, R. Sbiaa, A. Hirohata, H. Ohno, S. Fukami, and S. Piramanayagam, Spintronics based random access memory: a review, Materials Today **20**, 530 (2017).

[140] T. Heine, Transition metal chalcogenides: ultrathin inorganic materials with tunable electronic properties, Accounts of chemical research **48**, 65 (2015).

[141] C. Sergio *et al.*, Tuning Ising superconductivity with layer and spin–orbit coupling in two-dimensional transition-metal dichalcogenides, Nature Communications **9**, 1 (2018).

[142] N. Zibouche, A. Kuc, J. Musfeldt, and T. Heine, Transition-metal dichalcogenides for spintronic applications, Annalen der Physik **526**, 395 (2014).

[143] J. C. Koziar and D. O. Cowan, Photochemical heavy-atom effects, Accounts of chemical research **11**, 334 (1978).

[144] M.-H. Chiu *et al.*, Determination of band alignment in the single-layer $MoS_2/WSe_2$ heterojunction, Nature Communications **6**, 1 (2015).

[145] M. H. Chiu, W. H. Tseng, H. L. Tang, Y. H. Chang, C. H. Chen, W. T. Hsu, W. H. Chang, C. I. Wu, and L. J. Li, Band alignment of 2D transition metal dichalcogenide heterojunctions, Advanced Functional Materials **27**, 1603756 (2017).

[146] Y. Lv, Q. Tong, Y. Liu, L. Li, S. Chang, W. Zhu, C. Jiang, and L. Liao, Band-offset degradation in van der Waals heterojunctions, Physical Review Applied **12**, 044064 (2019).

[147] I. Grigorieva and A. Firsov, Electric field effect in atomically thin carbon films, Science **306**, 666 (2004).

[148] H. Wang, T. Kubo, J. Nakazaki, T. Kinoshita, and H. Segawa, PbS-quantum-dot-based heterojunction solar cells utilizing ZnO nanowires for high external quantum efficiency in the near-infrared region, The Journal of Physical Chemistry Letters **4**, 2455 (2013).

[149] G. Bai, Z. Yang, H. Lin, W. Jie, and J. Hao, Lanthanide Yb/Er co-doped semiconductor layered $WSe_2$ nanosheets with near-infrared luminescence at telecommunication wavelengths, Nanoscale **10**, 9261 (2018).

[150] H. P. Komsa and A. V. Krasheninnikov, Two-dimensional transition metal dichalcogenide alloys: Stability and electronic properties, Journal of Physical Chemistry Letters **3**, 3652 (2012).

[151] J. A. Hachtel, S. Susarla, V. Kochat, C. Tiwary, P. Ajayan, and J. C. Idrobo, Directly Identifying Phase Segregation in 2D Quaternary Alloys, Microscopy and Microanalysis **23**, 1438 (2017).

[152] J. Kang and J. Li, in *$MoS_2$: Materials, Physics, and Devices*, edited by Z. M. Wang (Springer, 2014), pp. 77.

[153] D. Rai, T. V. Vu, A. Laref, M. A. Hossain, E. Haque, S. Ahmad, R. Khenata, and R. Thapa, Electronic properties and low lattice thermal conductivity ($\kappa_l$) of mono-layer (ML) $MoS_2$: FP-LAPW incorporated with spin–orbit coupling (SOC), RSC Advances **10**, 18830 (2020).

[154] G. Ding, Y. Hu, D. Li, and X. Wang, A comparative study of thermoelectric properties between bulk and monolayer SnSe, Results in Physics **15**, 102631 (2019).

[155] S. Kumar and U. Schwingenschlögl, Thermoelectric performance of functionalized $Sc_2C$ MXenes, Physical Review B **94**, 035405 (2016).

[156] S. N. Sadeghi, M. Zebarjadi, and K. Esfarjani, Non-linear enhancement of thermoelectric performance of a $TiSe_2$ monolayer due to tensile strain, from first-principles calculations, Journal of Materials Chemistry C **7**, 7308 (2019).

[157] X. Zhao, C. Yao, T. Liu, J. C. Hamill Jr, G. O. Ngongang Ndjawa, G. Cheng, N. Yao, H. Meng, and Y. L. Loo, Extending the photovoltaic response of perovskite solar cells into the near-infrared with a narrow-bandgap organic semiconductor, Advanced Materials **31**, 1904494 (2019).

[158] N. T. Hung, A. R. Nugraha, E. H. Hasdeo, M. S. Dresselhaus, and R. Saito, Diameter dependence of thermoelectric power of semiconducting carbon nanotubes, Physical Review B **92**, 165426 (2015).

[159] F. A. Rasmussen and K. S. Thygesen, Computational 2D materials database: Electronic structure of transition-metal dichalcogenides and oxides, The Journal of Physical Chemistry C **119**, 13169 (2015).